\documentclass[aoas,preprint]{imsart}

\RequirePackage{amsthm,amsmath,amsfonts,amssymb}
\RequirePackage[authoryear]{natbib}
\RequirePackage[colorlinks,citecolor=blue,urlcolor=blue]{hyperref}
\RequirePackage{graphicx}
\renewcommand{\P}{\mbox{$\mathbb{P}$}}
\newcommand{\E}{\mbox{$\mathbb{E}$}}
\usepackage{subcaption}
\usepackage{float}
\usepackage{booktabs}
\usepackage{mdframed}
\usepackage{xr}
\startlocaldefs

\newtheorem{theorem}{Theorem}[section]

\theoremstyle{remark}
\newtheorem*{Remark}{Remark}
\newtheorem{method}{Method}[section]

\let\hat\widehat
\let\tilde\widetilde

\endlocaldefs

\begin{document}

\begin{frontmatter}
\title{Model-Independent Detection of New Physics Signals Using Interpretable Semi-Supervised Classifier Tests}
\runtitle{Signal Detection Using Interpretable Semi-Supervised Classifier Tests}

\begin{aug}
\author[A]{\fnms{Purvasha} \snm{Chakravarti}\ead[label=e1]{p.chakravarti@ucl.ac.uk}},
\author[B]{\fnms{Mikael} \snm{Kuusela}\ead[label=e2,mark]{mkuusela@andrew.cmu.edu}}
\author[C]{\fnms{Jing} \snm{Lei}\ead[label=e4,mark]{jinglei@stat.cmu.edu}}
\and
\author[B]{\fnms{Larry} \snm{Wasserman}\ead[label=e3,mark]{larry@stat.cmu.edu}}
\address[A]{Department of Statistical Science,
University College London,
\printead{e1}
}
\address[B]{Department of Statistics \& Data Science and NSF AI Planning Institute for Data-Driven Discovery in Physics,
Carnegie Mellon University,
\printead{e2,e3}}
\address[C]{Department of Statistics \& Data Science,
Carnegie Mellon University,
\printead{e4}
}
\end{aug}

\begin{abstract}
A central goal in experimental high energy physics is to detect new physics signals that are not explained by known physics. In this paper, we aim
to search for new signals that appear as deviations from known
Standard Model physics in high-dimensional particle physics data.  To do this, we determine whether there is any statistically significant difference between the distribution of Standard Model background samples and the distribution of the experimental observations, which are a mixture of the background and a potential new signal. Traditionally, one also assumes access to a sample from a model for the hypothesized signal distribution. Here we instead investigate a model-independent method that does not make any assumptions about the signal and uses a semi-supervised classifier to detect the presence of the signal in the experimental data. We construct three test statistics using the classifier: an estimated likelihood ratio test (LRT) statistic, a test based on the area under the ROC curve (AUC), and a test based on the misclassification error (MCE).  Additionally, we propose a method for estimating the signal strength parameter and explore active subspace methods to interpret the proposed semi-supervised classifier in order to understand the properties of the detected signal. We also propose a Score test statistic that can be used in the model-dependent setting. We investigate the performance of the methods on a simulated data set related to the search for the Higgs boson at the Large Hadron Collider at CERN. We demonstrate that the semi-supervised tests have power competitive with the classical supervised methods for a well-specified signal, but much higher power for an unexpected signal which might be entirely missed by the supervised tests.
\end{abstract}

\begin{keyword}
\kwd{collective anomaly detection}
\kwd{active subspace}
\kwd{mixture proportion estimation}
\kwd{signal strength estimation}
\kwd{likelihood ratio test}
\kwd{high-dimensional two-sample testing}
\kwd{Large Hadron Collider}
\end{keyword}

\end{frontmatter}

\section{Introduction}

Statistical and machine learning tools have been extensively used over
the past few decades to answer fundamental questions
in particle physics \citep{bhat2011multivariate, behnke2013data}.
To answer these questions, one needs to experimentally
test the predictions of the Standard Model, which describes our
current understanding of elementary particles and their interactions.
For example, the empirical confirmation of the Higgs
boson at CERN in 2012 was an essential step towards its inclusion in the Standard
Model \citep{aad2012observation, chatrchyan2012observation}.

In this paper, we develop statistical tools to address the problem of searching 
for evidence of new particle physics phenomena in high-dimensional experimental data, 
which is beyond what is explained by the Standard Model (SM). 
The goal is to search for a \emph{signal}, which represents any anomalous phenomenon that is unexplained by the Standard Model. 
On the other hand, any event predicted by the Standard Model, including all the previously discovered rich, structured signals predicted by the model, comprise the \emph{background}. 
For example, for more recent searches, background would include Higgs signals, which were only recently the focus of attention.
The search is performed on the observed unlabelled data, called here the \emph{experimental} data, which are a mixture of the background and a potential new signal.

In experiments conducted with large particle accelerators such as the
Large Hadron Collider (LHC), 
the searches for new physics signals in the high-dimensional experimental data have
been primarily conducted using model-dependent
data analysis methods. These searches are generally structured as
likelihood ratio tests based on a model assumption for the specific new
signal \citep{cowan2011asymptotic, atlas2011collaborations}. 
Due to the high-dimensionality of the data, tests based on classifiers that are optimized to detect a particular hypothesized signal are preferred over density estimation or mixture model approaches. Specifically, tests based on supervised,
multivariate classification algorithms such as neural networks and
boosted decision trees have proved beneficial. The training samples for the
classifier are generated using Monte Carlo (MC) event generators, 
which enable sampling collision events from specific physics models.
The classifier output is then used to extract a signal enriched sample which is used to 
perform likelihood ratio tests for the detection of the signal
\citep{aad2012observation, chatrchyan2012observation}.

\begin{figure}[t]
  \centering
  \begin{subfigure}{.45\textwidth}
  \includegraphics[width=\linewidth]{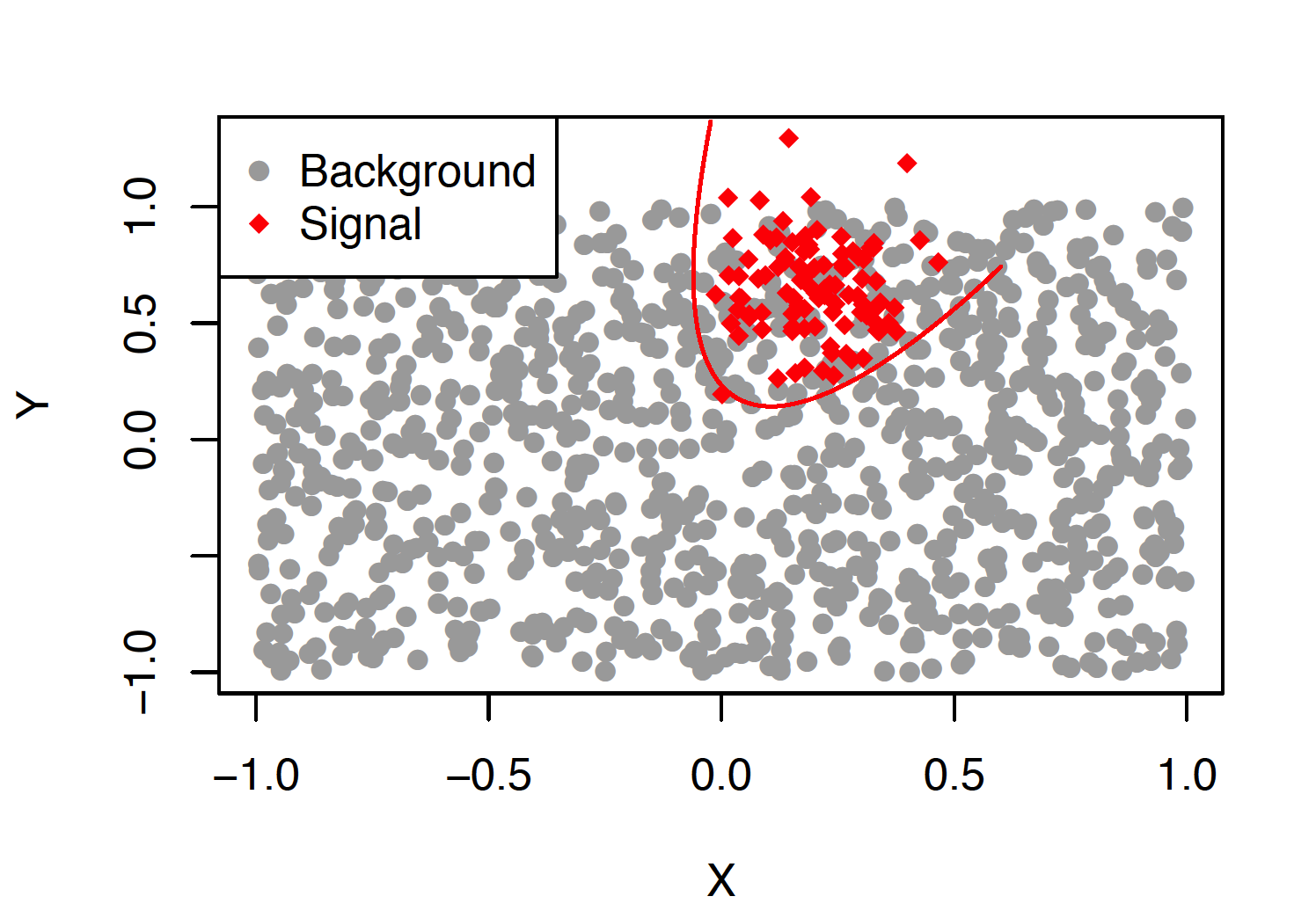}
  \caption{Training Samples}
  \end{subfigure}
  \begin{subfigure}{.45\textwidth}
  \includegraphics[width=\linewidth]{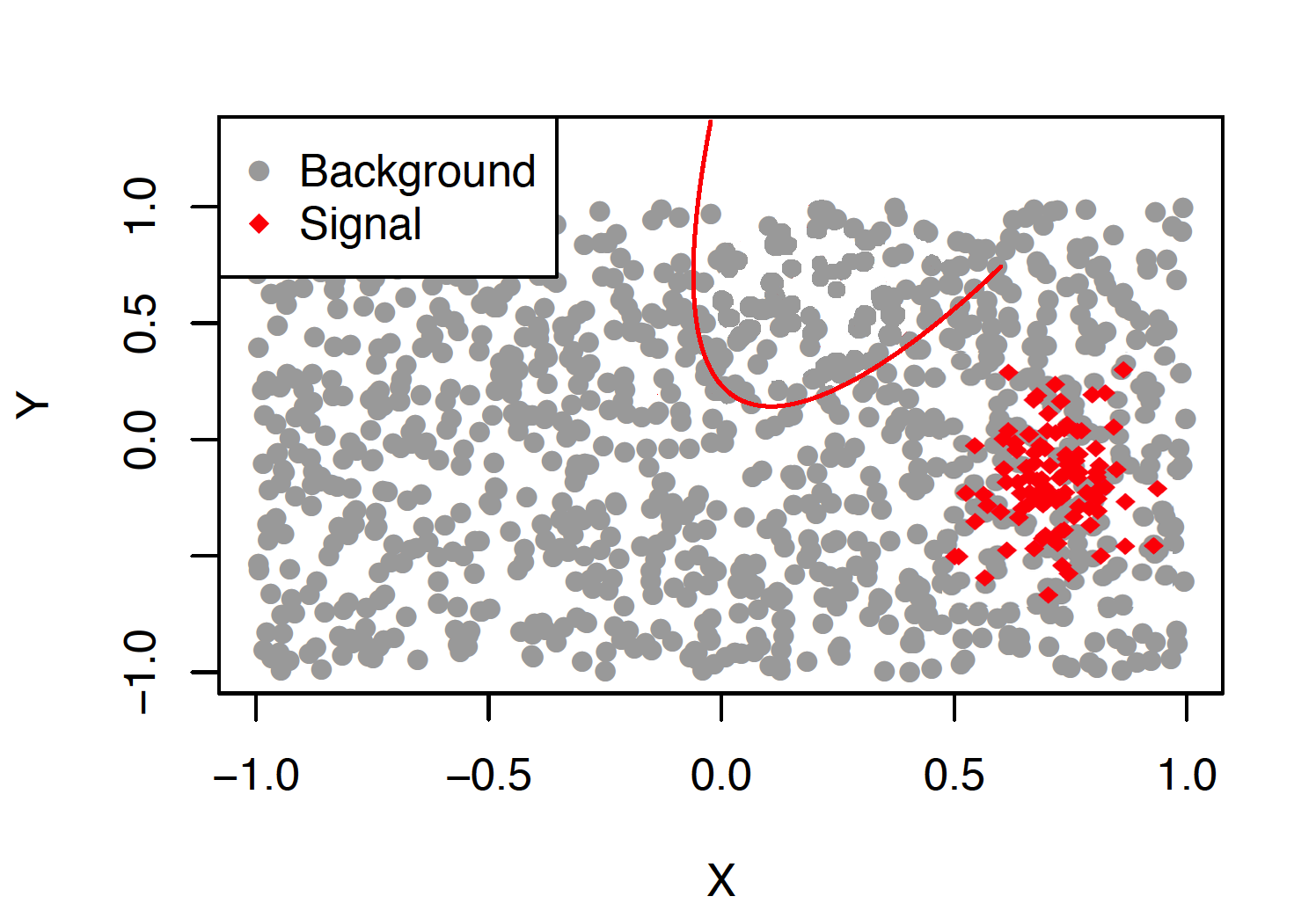}
  \caption{Experimental Samples}
  \end{subfigure}
\caption{Decision boundary using a supervised classifier to separate
  the signal (red rhombuses) from the background (grey). (a) The decision boundary of the
  classifier when trained on signal generated from the assumed signal
  model. (b) When the signal model is mis-specified, the classifier completely misses the actual signal data when used on the experimental samples. We consider a two-dimensional example here for illustration purposes only. The real data can be of much higher dimensionality. In the experiments considered in this paper, we have a 16-dimensional sample.}
\label{fig::superfails}
\end{figure}

But there are disadvantages to this
approach. First, this approach may have trouble finding novel,
unexpected deviations from the background model. 
Second, a search targeting one kind of new physics signal/scenario is not going to
be powerful for finding a different new physics signal/scenario. 
Figure~\ref{fig::superfails} illustrates
the problem. If a classifier is trained on training
signal data as shown in Figure~\ref{fig::superfails}(a), it gives the
classification boundary as shown. But what if the signal in the experimental data actually
looks like Figure~\ref{fig::superfails}(b)? Then the classifier ends
up misclassifying the signal as background. So an algorithm trained on
a mis-specified signal model might completely miss the actual signal. 
The two-dimensional example considered here 
is for illustration purposes only. 
In reality, the data lie in a high-dimensional space 
which further aggravates the problem.

In this paper, we study 
\emph{model-independent tests}, which do not assume a particular signal model, and compare them to
more traditional model-dependent tests 
for search of new physics signals.
We use data from
an event generator
for background events
together with observed
experimental data, 
which are a mixture of background and a potential signal.
But we do not use
data from signal simulations.
In other words,
we assume access to 
labeled background data and 
unlabeled experimental data,
where events may either have background or signal labels. 
(Note that in practice, systematic uncertainty
in the background makes the situation more complex than this. 
Please refer to Sections 2 and 7 for further discussion.) 
We then use a
semi-supervised approach that trains a classifier to differentiate the
background data from the experimental data.
Crucially, we do not assume availability of 
labelled signal data,
which differentiates this approach from model-dependent or supervised methods.

In particular, we make the following three main contributions: 
\begin{enumerate}
\item We propose and investigate several variants of classifier-based model-dependent and model-independent tests to detect a new signal in the experimental data. These tests involve four steps:
\begin{enumerate}
\item {\bf Training a probabilistic classifier} to differentiate between background events and potential new signal events in the model-dependent mode and to differentiate between background events and  experimental events in the model-independent mode.
\item {\bf Constructing classifier-based test statistics} using the output of the probabilistic classifier. We construct four different test statistics, using two different strategies, and compare them. 
\item {\bf Estimating the null distributions} of the test statistics using asymptotic, nonparametric bootstrap, and permutation methods, and comparing them.
\item {\bf Testing} for the presence of a new signal using the test statistics and their estimated null distributions. We consider each combination of a test statistic and a method for estimating its null distribution as a separate test method.
\end{enumerate}

\item We develop a method for {\bf estimating the signal strength} in the experimental data, which is a challenging problem in model-independent searches 
since we do not have information about the signal model 
to characterize the signal.
\item We develop {\bf active subspace methods} to interpret the signal detected by the semi-supervised classifier in the high-dimensional space.
\end{enumerate}

\begin{figure}[H]
\includegraphics[width = 0.96\linewidth]{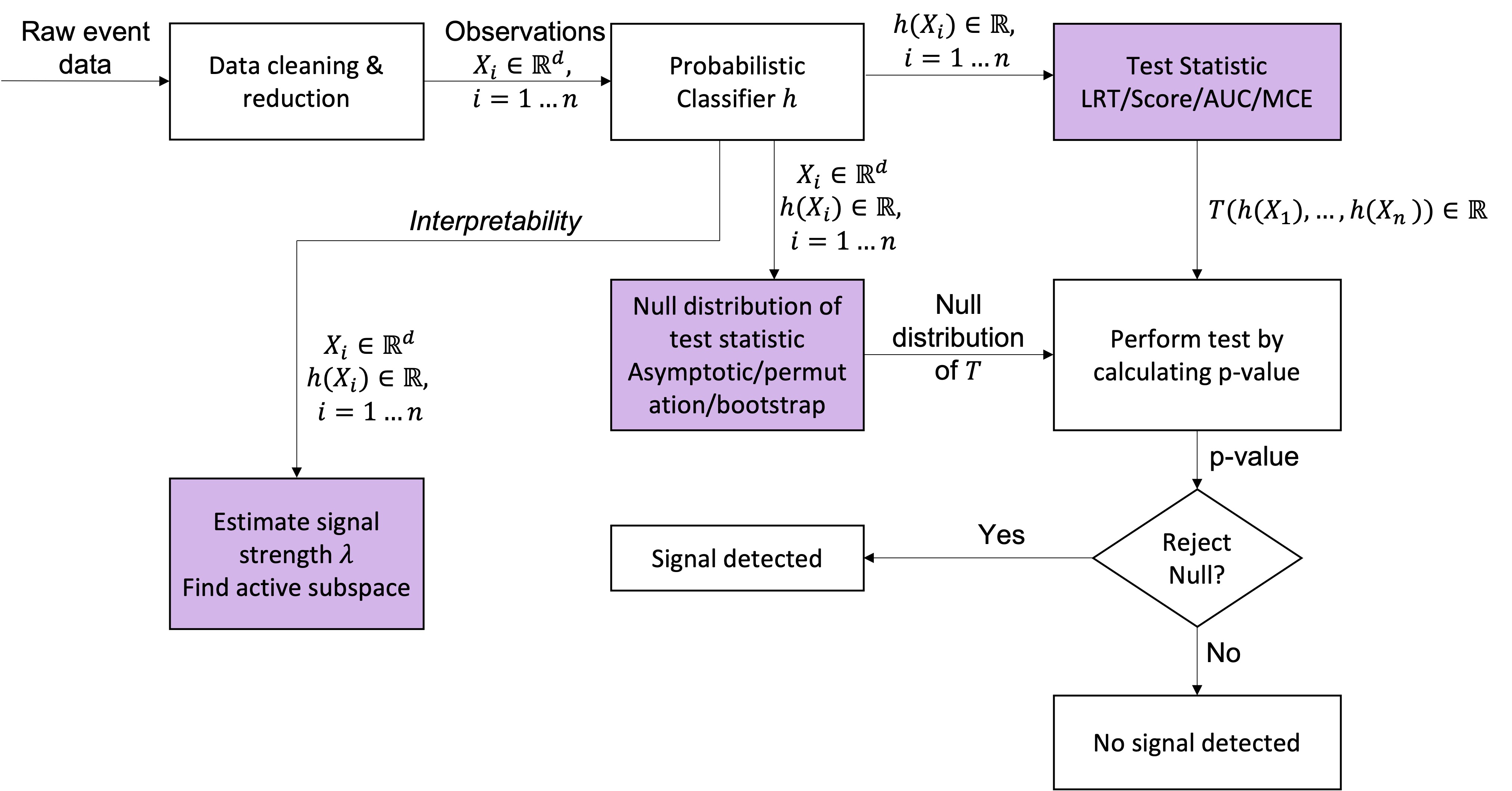}
\caption{Data flow diagram presenting the signal detection approach, including the interpretability of the trained classifier, from raw data to concluding whether there is any signal in the data. The colored boxes represent our contributions presented in this paper. The raw data and data reduction steps are presented, e.g., in \cite{pmlr-v42-cowa14}.}
\label{fig::dataflowdiag}
\end{figure}

The data flow diagram presented in Figure~\ref{fig::dataflowdiag}, illustrates the signal detection approach in four stages: training the probabilistic classifier, computing the test statistics, estimating the null distribution of the test statistics and finally performing the test. There is also the stage where we use the fitted classifier to estimate the signal strength and interpret the classifier using active subspace methods. The colored boxes in the diagram represent our contributions in this paper, namely introducing the different test statistics, introducing methods to estimate the null distribution, estimating the signal strength, and interpretability via active subspace methods. The diagram also demonstrates how the dimensionality of the data reduces from observations (16-D in our case) to the classifier outputs (1-D) which are then used to find the scalar test statistics. Below we detail selected parts of our approach.

{\bf Training a probabilistic classifier.} The approach we take is one introduced by particle physicists in the model-dependent mode, where we first train a probabilistic classifier ($h$) and then use its output to construct the test statistics. In the model-dependent (MD) mode, the classifier ($h$) finds differences between the properties of simulated background data and simulated signal data. The classifier ($h$) is then applied on the experimental data ($W_i$'s) and the output $(h(W_i))$ is used to construct the test statistics. In the model-independent (MI) mode, we do not make assumptions about the signal model, and hence do not have access to simulated signal data. So, the classifier instead is trained to find differences between the simulated background data and the experimental data, and if it is able to differentiate between them, it indicates the presence of a signal component in the experimental data.

{\bf Constructing classifier-based test statistics.} We propose two different strategies to construct test statistics from the probabilistic classifier. Using each of the two strategies, we then construct two test statistics. The first strategy, creating \emph{density ratio based test statistics},  takes advantage of the fact that for a probabilistic classifier, the output $h$ is the posterior binary class probability. Now, the posterior probability ($h(w)$) can  be written as a one-to-one function of the ratio ($\psi(w)$) of the probability densities of the two classes. Hence, the density ratio $\psi(w)$ can be estimated using the classifier output $h(w)$. Using these density ratio estimates, we create two classical test statistics -- the likelihood ratio test statistic (LRT) and the score statistic. Note that the two classes are background and signal in the MD mode and background and experimental in the MI mode. This strategy leverages the fact that
classifiers have been found to give
accurate estimates of density ratios
\citep{cranmer2015approximating}.

The second strategy is to construct \emph{classification performance based test statistics}. This strategy, which is only applicable in the MI setting, takes advantage of the fact that in the MI mode,  in the 
absence of a signal in the experimental data, a classifier should
not be able to differentiate the experimental data from the background
data, since they have the same distribution. This is analogous to a 
high-dimensional two-sample testing problem
where we compare the distributions of 
the background and the experimental data using a classifier \citep{friedman2004multivariate, kim2016classification, kim2019global}. In this case, the performance of the classifier is indicative of whether there is any signal in the data. 
We measure the performance in two ways, resulting in two test statistics -- the area under the ROC curve  (AUC) and the misclassification error (MCE). The reason we include these test statistics is because they are properties of the classifier itself in contrast to the LRT which is being estimated using the classifier. 

The four test statistics according to the two strategies are presented in Table~\ref{tbl::TestStatistics}. In the model-dependent mode we cannot use classification performance based test statistics since the classifier is trained on background and signal data. In the model-independent mode, we cannot use the score statistic. This is because the score statistic is a function of the density ratio of the \emph{signal} data and the background data, whereas, in the model-independent setting, we only have an estimate for the density ratio of the \emph{experimental} data and the background~data.

\begin{table}[H]
\caption{Test statistics considered for model-dependent (MD) and model-independent (MI) searches.``X" marks cells that do not have a corresponding test statistic.}
\label{tbl::TestStatistics}
\centering
\begin{tabular}{@{}lccccc@{}}
\cmidrule{2-6}
 & \multicolumn{5}{c}{\bf{\normalsize Test Statistics}} \\ \cmidrule{2-6} 
 & \multicolumn{2}{c}{\bf{Density Ratio Based}} & & \multicolumn{2}{c}{\bf{Classification Performance Based}} \\ \cmidrule{2-3} \cmidrule{5-6} 
 & \multicolumn{1}{c}{\bf{LRT}} & \multicolumn{1}{c}{\bf{Score}} & & \multicolumn{1}{c}{\bf{AUC}} & \bf{MCE} \\ 
 \cmidrule{2-3} \cmidrule{5-6} 
\multicolumn{1}{l}{\bf{Model-dependent}} & \multicolumn{1}{c}{MD-LRT} & \multicolumn{1}{c}{MD-Score} & & \multicolumn{1}{c}{X} & \multicolumn{1}{c}{X} \\ 
\multicolumn{1}{l}{\bf{Model-independent}} & \multicolumn{1}{c}{MI-LRT} & \multicolumn{1}{c}{X} & & \multicolumn{1}{c}{MI-AUC} & \multicolumn{1}{c}{MI-MCE} \\ \cmidrule(l){2-6} 
\end{tabular}
\end{table}

One of our model-independent tests based on the LRT statistic 
is similar to the test proposed by \cite{d2019learning} 
and \cite{d2021learning}. 
The other two model-independent test statistics are the AUC and MCE statistics.
To the best of our knowledge, these two are novel in this application. 
Though the test that uses the LRT  
is similar to the test proposed by  \cite{d2019learning} 
and \cite{d2021learning}, 
unlike them, we use the nonparametric bootstrap 
as well as permutation methods to estimate the null distribution, 
comparing the 
performance of the different re-sampling techniques. 
We also apply the tests in a higher-dimensional setting than them.
Additionally, since our test has a simple null versus 
a composite alternative, the LRT is not guaranteed to yield an optimal test in this setting. 
Hence, it is important to compare 
the performance of the different test statistics. 

{\bf Estimating the null distributions} and {\bf Testing.} We estimate the null distributions of the test statistics using four different approaches - asymptotic, nonparametric bootstrap, permutation, and slow permutation. The bootstrap and permutation approaches are resampling methods that utilize the fact that under the null, the experimental data and the background data have the same distribution and so, the data can be resampled from a combination of the two data sets and their labels can be permuted. The bootstrap and the regular permutation method split the data into training and test data sets and re-sample the test data sets only, whereas the slow permutation method re-samples the entire data set and re-trains the classifier on each re-sampled data set, which increases its computational complexity. A combination of a test statistic and a null distribution estimation method forms a test. In principle, every test statistic has a true null distribution, which these methods are trying to estimate. So, under the null, these methods should give similar distributions. However, it is important to note that the estimated null distribution is a function of the experimental data, and hence only when there is no signal in the experimental data (when null is true), it estimates the true null distribution. So, the power of the tests, when there is signal, is expected to be different for the different null estimation methods. Hence, it is important to compare the performance of the different null estimation methods for the different test statistics.

{\bf Estimating the signal strength.} 
Towards our second contribution of estimating the signal strength,
 the problem is particularly challenging in the MI mode, since the classifier differentiates between background and experimental data, and hence cannot directly identify signal events. 
We solve this by showing that  
estimating the signal strength is equivalent to 
estimating a monotone univariate density over the interval $[0, 1]$ at the right boundary at 1. 
In particular, we need to estimate the density of a statistic 
built by considering the quantile of a Neyman--Pearson-style likelihood ratio. 
Similar to the estimator given by \cite{storey2002direct}, 
which is a special case of a boundary density estimator
using histograms, 
we use histograms to estimate the density, 
and then use a Poisson regression \citep{nelder1972generalized} model 
of the histogram bin counts 
to estimate the density at the boundary point. 
The additional use of Poisson regression 
makes the boundary estimate more stable. 
We also find confidence intervals for the estimates using GLM regression intervals 
and three different bootstrap methods.

{\bf Active subspace methods.}
For our third contribution, 
to interpret the classifier and characterize the signal, 
we propose using active subspace methods 
\citep{constantine2014active, constantine2015active} 
to explore which aspects of the covariates are the most informative for the test. 
Active subspaces have been used as a form of sensitivity analysis 
to quantify uncertainty in an output  \citep{constantine2015exploiting} and 
they have also been used to
analyze the internal structure and 
vulnerability of deep neural
networks \citep{cui2020active}. 
Similar to these works, 
in this paper, we propose active subspace methods 
to interpret the variability in the classifier output 
in terms of its inputs.
Our proposed methods identify  
the directions that capture the most variability 
in the gradient of the classifier. 
These directions are found using an eigen decomposition (PCA) 
of the gradients of the classifier surface at observed points. 
These directions give us information about 
the combinations of the covariates
 that most influence the classifier in 
 distinguishing the experimental data 
 from the background data, 
 hence giving us an idea of what the signal looks like.
 Importantly, this process helps us better understand the classifier that detects the signal,
 which otherwise would be a black box.

Alternative variable importance methods have been used
to understand the intrinsic predictiveness potential of the covariates
 \citep{van2006statistical, lei2018distribution, williamson2020unified} 
 which identify the individual covariates 
 that play an important role in the classifier.
Variable importance has been addressed
extensively for random forests \citep{breiman2001random, ishwaran2007variable, strobl2008conditional, gromping2009variable} and
neural networks \citep{bach2015pixel, shrikumar2017learning, sundararajan2017axiomatic}.
But contrary to the active subspace methods, 
these methods concentrate on finding the variables that are 
individually the most important and potentially miss detecting 
combinations of covariates that might have more predictive power.

It is worth noting that the methods presented in this paper can also be applied more generally. In this paper, we demonstrate an application of the methods to the search of new phenomena in particle physics, but the methods presented here are applicable beyond particle physics. In general, these methods can be applied whenever there is a need to search for, detect or characterize collective anomalies in a high-dimensional space, where a single data point is not necessarily anomalous but a collection of data points is. For example, these methods could potentially be used to find anomalous weather changes in high-dimensional climate science data sets, where individual weather events might themselves not be anomalous but a collection of certain types of weather events over a period of time might be. Potential further applications range from engineering to medicine and other areas of the physical sciences.

The rest of this paper is structured as follows. In  the   following  section, 
we describe the role of model-independent methods in the search for new physics phenomena
and discuss some existing literature on it. 
In Section~\ref{sec::methods},   we 
introduce  the   problem  setup   mathematically.  We describe
the MD methods  in Section~\ref{sec::supervised} and  the
proposed MI  methods  in
Section~\ref{sec::semisupervised}.  
We then discuss methods to 
estimate the signal strength in the experimental data 
in Section~\ref{sec::lambdaEst}. 
In   Section~\ref{sec::active},  we  describe  active
subspace methods  to understand the subspace  affecting the classifier
the most, leading to an understanding of the detected signal. 
Finally, in
Section~\ref{sec::higgs},  we  demonstrate   the  performance  of  the
proposed MI methods and compare them to the MD methods using 
publicly available simulated data from a machine learning data challenge 
that includes events corresponding to the SM Higgs boson signal. 
We include concluding remarks and possible future directions of research in Section~\ref{sec::conclusion}. We also provide exploratory analysis 
of the Higgs boson data set that is used to perform the experiments in 
Section~\ref{sec::higgs} and include some other experimental details 
in the Supplementary Material \citep{supplementary}. 
We defer the proofs of the theorems and some of the proposed algorithms to the Supplementary Material as well.

\section{Overview of Model-Independent Searches of New Physics}
\label{sec::discussion}

In particle physics, before any search analysis is performed, 
the experimental data is typically filtered for specific types of final-state particles. Different choices here specify different search channels. 
For example, the search could be restricted to a decay channel where four muons are observed. 
The data set would then contain kinematic and other features of these particles observed using the particle detector, which dictates the dimensionality of the data. 
The data set is further restricted by selection cuts on these features that are used to remove data points that lie in regions of the phase space that are believed to be signal-free. 
For example, one might choose only those events where the particles have large transverse momenta (larger than some threshold). 
These cuts are generally performed to increase the signal-to-background ratio in the final experimental data that is considered in the analysis.
Even after the selection cuts, a typical data set that needs to be analyzed may have a sample size ranging from hundreds of thousands of events to million or billions of events.
All of the methods presented in this paper can handle such large data sets (as demonstrated in Section~\ref{sec::higgs}). 
In the experiments performed in this paper, we use a 16-dimensional data set which is also typical for the searches, although the dimensionality can also be much larger for complex search channels or if low-level detector information is used.

Most model-independent approaches find the signal in experimental data 
by comparing it to a reference background dataset, 
which is an essential requirement for these strategies. 
Hence, it is necessary for these approaches to have a trustable reference dataset. 
As discussed in \cite{d2022learning}, 
it is conceptually unavoidable for these model-independent approaches 
to require a reference background dataset,  
since they search for ``new'' phenomena, and 
hence require a necessary notion of ``old'' phenomena. 
The required reference background datasets are usually generated using Monte Carlo event generators, which sample collision events consistent with the equations of the Standard Model of particle physics. 
Some recent approaches, such as ANODE \citep{nachman2020anomaly} and CWoLa (Classification Without Labels) Bump Hunting \citep{collins2018anomaly, collins2019extending}, 
estimate the background from the data itself, 
which requires assumptions on the signal region and 
access to data not in the signal region. 
Here we do not make any assumptions on the signal region, 
so we use Monte Carlo simulations from the Standard Model as the reference background dataset. 
However, it is good to note that model-independent approaches are not necessarily restricted to Monte Carlo backgrounds, as demonstrated by ANODE and CWoLa.

Even though these model-independent approaches are dependent on 
the availability of a reliable, trustable background dataset to find new phenomena, 
the key advantage of model-independent approaches is that they can detect
discrepancies between the background data and the experimental data
irrespective of the distribution of the signal events. Such capability can be essential for ensuring the maximal reach of the LHC physics program. In the case that
a discrepancy is found, it should be
investigated further in order to understand if it results from (a) an
inaccurate background MC generator, (b) a particle detector defect
or a lack of understanding of the detector, or (c) a previously
unknown physics process. 
The active subspace methods proposed in this paper may provide a
preliminary indication in this process. A model-independent search indicating a significant discrepancy can also guide the selection of new model-dependent analyses to further investigate the nature of the discrepancy.

It is important to note here that there is no one ``right'' approach to finding new physics signals. 
The different model-dependent and model-independent approaches are 
complementary and not superior to each other.  
For example, when a reliable signal model is available, 
model-dependent approaches should yield greater power 
since they use the information available about the signal model to optimize the test. It is also good to note that there is no black-and-white distinction between model-independent and model-dependent methods. Rather, there is a continuum of methods that vary in terms of the strength of the assumptions involved. Any ``model-independent'' method at minimum needs to make assumptions about the background. Furthermore, to perform the model-independent search, one would in practice use some amount of physics knowledge to choose a particular decay channel (say, 4 muons) and to impose further cuts within that channel (e.g., to choose events where the particles have large transverse momenta) to narrow down the search to a part of the phase space that is believed to be fertile for new signals and where the background can be modeled sufficiently well. One might then be willing to make further assumptions about the location or shape of the signal region, and so on. At the far end of this spectrum are fully ``model-dependent'' methods that assume a specific signal model.

Additionally, both model-independent and model-dependent methods are affected by
systematic uncertainties in the background \citep{cranmer2015practical, Lyons_2018, dorigo2020dealing}.
Background systematics in high-energy physics are typically parameterized using nuisance parameters that affect, among other things, the rate and shape of the background samples. In low-dimensional situations, a standard approach is to handle these nuisance parameters using likelihood profiling \citep{cranmer2015practical}, while recent work has focused on incorporating nuisance parameters into classifier training for high-dimensional data. Training a classifier on simulated data generated using specific values for the nuisance parameters may not be optimal for handling background data generated using other values of the nuisance parameters. Even if the classifiers are trained on data generated using the most likely values of the nuisance parameters, and their effect is accounted for in the calibration, the power of the methods to detect new signals may be affected \citep{dorigo2020dealing}. Some classifier-based model-dependent approaches have been recently suggested to handle the nuisance parameters. For example, \cite{Ghosh:2021roe} profile the nuisance parameters by constructing classifiers that are explicitly dependent on the different values of the nuisance parameters. Reviews of model-dependent approaches that deal with nuisance parameters in classifier-based searches can be found in \cite{10.21468/SciPostPhys.8.6.090} and \cite{dorigo2020dealing}. Overall, dealing with systematic uncertainties in the background in the model-independent setting is a challenging problem that affects any model-independent method and whose detailed treatment is beyond the scope of this paper. However, we note there is no reason to believe that it would not be possible to extend existing approaches for handling background systematics from the model-dependent setting into the model-independent case. \cite{d2022learning} is, to the best of our knowledge, a first contribution toward this important goal.

Below, we present a non-exhaustive review of existing literature on model-independent approaches and other related work.

\subsection{Related Literature}
\label{sec::relatedliterature}

Due to their advantages, model-independent approaches have been used
for new physics searches \citep{knutesonPHD, CDF:2007iou, Soha:2008vd} at the Tevatron \citep{Choudalakis:2008pr, aaltonen2009global,
 bertram2012model}, HERA \citep{aktas2004general}, and the LHC
\citep{cms2017music, aaboud2019strategy, cms2020music}. 
These methods typically
compare a large set of binned distributions to the prediction from the
background Monte Carlo simulation, in search for bins in the
experimental data that exhibit a deviation larger than some predefined
threshold. For example, the approach of \cite{aaboud2019strategy},
 employed by the ATLAS experiment, 
 uses a (quasi\nobreakdash-)model-independent method that considers
some generic features of the potential new physics signals. 
These
approaches have two limitations: 
(a) they do not consider the multivariate dependency
structures between the variables in the data and 
(b) they might miss
certain signals that do not show a localized excess in one of the
studied distributions. 

More recent approaches like  CWoLa (Classification Without Labels) Bump Hunting \citep{collins2018anomaly, collins2019extending}, 
ANODE  \citep{nachman2020anomaly}, 
SALAD \citep{andreassen2020simulation} 
and simulation augmented CWoLa (SA-CWoLa)  \citep{kasieczka2021lhc} are also based on  searching for 
anomalies by assuming that the signal is localized in a single feature.   
These approaches use this weak assumption about the signal which causes them to have the second limitation mentioned above, namely they might miss signals that do not show a localized excess along one of the features. 
Some of the algorithms additionally assume that the signal’s distribution for the other features is
independent of the selected single feature that is being scanned.  
\cite{kasieczka2021lhc} describes a variety of current methods for signal detection along with results on the LHC Olympics 2020 datasets.

A different approach that uses a semi-supervised nonparametric clustering algorithm
 is presented by \cite{casa2018nonparametric}. 
They assume that in high energy physics, 
a new particle manifests itself as a significant peak
emerging from the background process. They use nonparametric modal
clustering to search for a signal that is expected to emerge as a bump
in the background distribution. 
BuHuLaSpa and Bump Hunter methods 
presented in \cite{kasieczka2021lhc} also use similar bump hunting 
ideas to find the signal. UCluster presented in  \cite{kasieczka2021lhc} 
 uses clustering to find the localized signal. 
These methods suffer from the second
problem mentioned above: they can only find localized signals.

Model-independent semi-supervised searches were also proposed by
\cite{kuusela2012semi} and \cite{vatanen2012semi} who use multivariate
Gaussian mixture models to estimate the densities of the background
and the experimental data.  
They test 
for the significance of the
additional Gaussian components which quantify the anomalous
contribution. The drawback of this method is that Gaussian mixture
models are very difficult to fit in a high-dimensional
setting. Additionally, since the signal strength is typically very
small, the quality of the fit influences the power of the test in
detecting the signal.

These drawbacks of the mixture modeling methods 
 along with the fact  (as mentioned before) that  
 classification algorithms have demonstrated 
excellent performance in detecting signals in the model-dependent
approaches, especially in the high-dimensional setting,  
motivates us to use classifiers instead of 
mixture modeling to find the
deviations of the experimental data from the background. 

The problem of 
comparing the distributions of 
the background and the experimental data
is also analogous to a high-dimensional two-sample testing problem 
\citep{kim2016classification, kim2019global}, where 
the signal events appear as a collective anomaly \citep{chandola2009anomaly} 
in a cluster close to each other. 
Hence, we also compare the methods proposed in this paper to   
nearest-neighbor    two-sample    tests     introduced    in
\cite{schilling1986multivariate} and \cite{henze1988multivariate} 
for a sub-sample of the data. 
In the experiments, we observe that methods that use 
a semi-supervised classifier have much higher power 
to detect the signal than the nearest-neighbor two-sample tests.

\section{Classifier-Based Tests for Signal Detection}
\label{sec::methods}

In this section, we introduce 
both model-dependent and model-independent approaches 
to signal detection in the experimental particle physics data. 
Both  approaches use background data from 
MC event generators for background events based on the Standard Model, 
together with experimental data, 
which are a mixture of background and a potential signal. 
First, we discuss the model-dependent approach that 
additionally assumes access to signal data 
that are generated using MC event generators 
for a hypothesized signal model. 
Then we describe the model-independent approach
 that does not assume access to the signal data.
We now introduce formal notation to discuss the two different approaches.

The background, signal and experimental data are samples from
Poisson point processes \citep{cranmer2015practical, reiss2012course}.
We condition on the sample sizes of the individual datasets
so that the data in all three of the cases
may be treated as independent samples
from a density (i.e., from a binomial point process).
Specifically, we have three datasets:
\begin{align}
\nonumber
\text{Background:} &\quad \mathcal{X} = \{ X_1,\ldots, X_{m_b} \}, &\quad X_i \sim p_b\\
\label{eq::pw} 
\text{Signal:} &\quad \mathcal{Y} = \{Y_1,\ldots, Y_{m_s}\}, &\quad Y_i \sim p_s\\
\nonumber
\text{Experimental:} &\quad \mathcal{W} = \{W_1,\ldots, W_n\}, &\quad W_i \sim q = (1-\lambda)p_b + \lambda p_s,
\end{align}
where $p_b, p_s$ are the densities of the background and signal data respectively,
 $q$ is the density of the experimental data 
 and $\lambda$ is a scalar parameter representing the signal strength.

The likelihood function for $\lambda$ given the experimental data ($\mathcal{W}$) 
(treating $p_b$ and $p_s$ as known for the moment)
is
\begin{equation}
{\cal L}(\lambda) =
\prod_i ( (1-\lambda)p_b(W_i) + \lambda p_s(W_i)).
\end{equation}
The null hypothesis ($H_0$) that there is no signal
corresponds to
$\lambda=0$. 
So the goal is to test $H_0: \lambda = 0$ 
versus $H_1: \lambda > 0$.
We additionally have the likelihood ratio:
\begin{equation}
\frac{ {\cal L}(\lambda)}{ {\cal L}(0)} =
\prod_i ( 1-\lambda + \lambda \psi(W_i))
\label{eqn::LR}
\end{equation}
where
$\psi(w) = p_s(w)/p_b(w)$.
Note that the function $\psi$
can be seen as an infinite dimensional nuisance parameter.

In the idealized case
where $p_b$ and $p_s$ are known,
we could use the usual likelihood ratio test (LRT) statistic
\begin{equation}
T = - 2 \log ( {\cal L}(0)/{\cal L}(\hat\lambda)) 
= 2 \sum_i \log (1-\hat\lambda + \hat\lambda \psi(W_i)),
\label{eqn::supervisedLRT}
\end{equation}
where $\hat\lambda$ is the maximum likelihood estimator (MLE) of $\lambda$
based on the experimental data $\mathcal{W}$.
Alternatively, we could use the score test statistic
\begin{equation}
\mathcal{T} = \frac{1}{n}\sum_i (\psi(W_i) -1).
\label{eqn::supervisedScore}
\end{equation}
In practice,
the densities $p_b$ and $p_s$ are unknown.
The two different approaches described in this section
show how a classifier can be used to estimate the desired statistics directly. 
We prefer estimating the density ratio required for the desired statistics directly 
instead of taking the ratio of the estimated densities. 
This is due to the high-dimensionality of the data 
which makes estimating the high-dimensional density with limited data very difficult.

\subsection{The Model-Dependent (Supervised) Case}
\label{sec::supervised}

In this case,
we make use of the signal data $\mathcal{Y}$, where
$Y_1,\ldots, Y_{m_s}  \sim p_s$ are
 generated using a MC event generator 
for a hypothized signal model. 
Such approach is standard 
in most new physics searches at the LHC, 
and serves here as a point of comparison 
against the model-independent methods.
The strategy underlying our implementation of this approach is to
use a classifier to
estimate $\psi = p_s/p_b$ and then use the 
LRT or the score test with the estimated $\psi$.\footnote{In most LHC analyses, $\psi$ 
is currently used to extract 
a signal-enriched subset of the data 
which is then used in a low-dimensional parametric fit 
to form a test statistic \citep{radovic2018machine}. 
Here we use instead the $\psi$-based 
high-dimensional LRT or score test 
to provide an apples-to-apples comparison 
with the model-independent methods.}
Since $\psi$ is estimated,
we cannot rely on standard asymptotics to get
the null distribution.
Instead we use permutation or nonparametric bootstrap methods.

Before we train a classifier, we first combine the 
background and signal data into a single dataset
\[ \{Z_1, \ldots, Z_{m_b+m_s}\} = \mathcal{X} \cup \mathcal{Y} = \{X_1, \ldots, X_{m_b}, Y_1, \ldots, Y_{m_s}\} \]
and we define $S_i = 1$ if $Z_i$ is from the signal data $\mathcal{Y}$ 
and $S_i = 0$ otherwise. 
We treat
$(Z_i, S_i)$
as a sample
from a density $p(z, s)$ 
with
$\pi_0 := \P(S=1) = m_s/(m_b+m_s)$,
the probability of any sample being from the signal distribution. 

Let $h_0(z) := {\P}\left(S = 1| Z = z \right)$ denote 
the probability of a sample being from the signal distribution given $Z = z$.
Then using Bayes's rule,
\begin{equation}
h_0(z) = {\P}\left(S = 1| Z = z \right) 
= \frac{\pi_0 \psi(z)}{\pi_0 \psi(z) + (1-\pi_0)}.
\end{equation}
Inverting this,
\begin{equation}
\psi(z) =
\left(\frac{1-\pi_0}{\pi_0}\right)
\left(\frac{h_0(z)}{1-h_0(z)}\right).
\end{equation}
This leads to our estimate of $\psi$,
\begin{equation}
\widehat{\psi}(z) = 
\left(\frac{1-\pi_0}{\pi_0}\right)
\left(\frac{\hat h_0(z)}{1-\hat h_0(z)}\right), \label{eqn::psihat}
\end{equation}
where $\hat h_0(z)$
is a classifier that separates 
the signal data $\mathcal{Y}$ from the background data $\mathcal{X}$.
In this paper, we take
$\hat h_0$ to be a random forest, 
but in principle, any classifier can be used.

To use $\widehat{\psi}(z)$ to calculate the LRT statistic in (\ref{eqn::supervisedLRT}),
 we estimate $\lambda$ using its MLE as follows.
We can write
the likelihood of the experimental data as
\begin{equation}
{\cal L}(\lambda) =
\prod_i p_b(W_i) \times
\prod_i (1-\lambda + \lambda \psi(W_i)).
\end{equation}
The first term does not involve $\lambda$ so we can ignore it.
Hence,
\begin{equation}
{\cal L}(\lambda) \propto
\prod_i (1-\lambda + \lambda \psi(W_i)).
\end{equation}
We then define the estimated likelihood
\begin{equation}
\hat{\cal L}(\lambda)  \propto
\prod_i (1-\lambda + \lambda \hat\psi(W_i))
\end{equation}
and we define
$\hat\lambda$ to be the maximizer of
$\hat{\cal L}(\lambda)$. 
We can now use $\hat\lambda$ and $ \hat\psi$ to estimate
the LRT and the score test statistics 
as given in (\ref{eqn::supervisedLRT}) and (\ref{eqn::supervisedScore}).

To see the effect of maximizing the estimated likelihood using the estimated $\psi$
instead of maximizing the actual likelihood,
let $\ell(\lambda) = \log {\cal L}(\lambda)$,
let $\hat\ell(\lambda) = \log \hat{\cal L}(\lambda)$,
and note that, for small $\lambda$
\begin{equation}
\ell(\lambda) =
\lambda \sum_i (\psi(W_i)-1) - 
\frac{\lambda^2}{2} \sum_i (\psi(W_i)-1)^2 + o_P(\lambda^2) + C,
\end{equation}
where $C = \sum_i \log(p_b(W_i))$ is just a constant and can be ignored.
A similar relation holds for $\hat \ell(\lambda) $ and $\hat \psi$.
So
\begin{equation}
\frac{1}{n}\hat\ell(\lambda)-
\frac{1}{n}\ell(\lambda)=
\frac{\lambda}{n}\sum_i (\hat\psi(W_i)-\psi(W_i)) + o_P(\lambda)
\end{equation}
which shows how the accuracy of the
classifier affects the log-likelihood.
The maximizer $\tilde\lambda$ of $\ell(\lambda)$ and
the maximizer $\hat\lambda$ of $\hat\ell(\lambda)$
are
\begin{equation}
\tilde\lambda =
\left[\frac{\sum_i (\psi(W_i)-1)}{\sum_i (\psi(W_i)-1)^2}\right]_+ + o_P(\lambda),\ \ \ \ \ 
\hat\lambda =
\left[\frac{\sum_i (\hat\psi(W_i)-1)}{\sum_i (\hat\psi(W_i)-1)^2}\right]_+ + o_P(\lambda).
\end{equation}
Hence,
$\hat\lambda - \tilde\lambda = O_P(\frac{1}{n}\sum_i (\hat\psi(W_i)-\psi(W_i)))$,
emphasizing again the importance of an accurate classifier. 
Note that in practice, instead of using the above approximation, 
we evaluate the MLE of $\lambda$ by performing a grid search on $[0, 1]$.

As $n \to \infty$, the usual likelihood ratio test statistic 
\citep{ghosh1984asymptotic, bohning1994distribution},
\begin{equation}
T =  2 \sum_i \log\left( (1-\hat\lambda) + \hat\lambda \psi(W_i)\right)  
\overset{d}{\rightsquigarrow} \frac{1}{2} \delta_0 + \frac{1}{2} \chi^2_1, 
\label{eqn::superLambda}
\end{equation}
where $\delta_0$ is a degenerate distribution at $0$. 
But when $\hat\psi$ is substituted for $\psi$,
the asymptotic distribution is unknown
so the null distribution needs to be estimated
by simulating from
the background model.
If the available background data is limited,
we can use permutation or nonparametric bootstrap methods;
see Section~\ref{sec::testperformance}.

Similar remarks apply to the score statistic
$\mathcal{T} = n^{-1}\sum_i (\psi(W_i)-1)$
and its estimated version
$\hat{\mathcal{T}} = n^{-1}\sum_i (\hat\psi(W_i)-1)$.
There are conditions under which
$\hat{\mathcal{T}}$ has a tractable distribution.
Suppose that $\hat\psi$ is estimated on part of the data
and $\hat{\mathcal{T}}$ on another.
Now
\begin{equation}
\sqrt{n} \hat{\mathcal{T}} = \sqrt{n} \mathcal{T} + \sqrt{n}\left(\frac{1}{n} \sum_i \left(\hat\psi(W_i)-\psi(W_i)\right)\right).
\end{equation}
The first term converges to
$N(0,\sigma^2)$ where
$\sigma^2 = \mathbb{E}_{p_b}[(\psi(W)-1)^2]$.
If $\psi$ is in a H\"{o}lder class of smoothness index $\beta$ and
$\beta > d/2$, then it can be shown that
$\frac{1}{n} \sum_i (\hat\psi(W_i)-\psi(W_i)) = O_P(n^{-2\beta/(2\beta+d)})$,
 where $d$ is the dimension of the $W_i$'s. 
Hence, if $\beta > d/2$,
the second term is negligible so that
$\sqrt{n}\hat{\mathcal{T}} \rightsquigarrow N(0,\sigma^2)$.
However, we have found that the Normal
approximation is poor in practice
and instead we use permutation and bootstrap methods
to approximate the null distribution.
Similar to the LRT, 
the null distribution can also be estimated 
by simulating from the background model.

To use nonparametric bootstrap or permutation methods, 
we randomly split the available background data $\mathcal{X}$
into two sets $\mathcal{X}_1$ and $\mathcal{X}_2$.
We use the first set $\mathcal{X}_1$ 
along with the signal data $\mathcal{Y}$ 
to train the classifier $\hat{h_0}$.
We use the second set $\mathcal{X}_2$ 
along with the experimental data $\mathcal{W}$ to 
approximate the null distributions of the LRT and the score statistics.
Note that, under the null $H_0: \lambda = 0$, 
the experimental data $\mathcal{W}$ and 
background data $\mathcal{X}_2$ have the same distribution $p_b$.
In the case of nonparametric bootstrap, 
we approximate the null distributions by 
repeatedly sampling with replacement 
from $\mathcal{X}_2 \cup \mathcal{W}$,
and computing the test statistics.
For the permutation test, 
we do the same using sampling without replacement.

\subsection{Model-Independent (Semi-Supervised) Case}
\label{sec::semisupervised}

In this case, we assume that we do not have access to (or do not
completely trust) the signal training sample 
$\mathcal{Y} = \{Y_1,\ldots, Y_{m_s}\}$. 
So the data available are
$\mathcal{X} = \{X_1,\ldots, X_{m_b}\}$
and
$\mathcal{W} = \{W_1,\ldots, W_n\},$ where 
$X_i  \sim p_b$ and $W_i  \sim q = (1-\lambda)p_b + \lambda p_s$, 
with $p_b$, $p_s$ and $\lambda$ unknown.
We want to test
$H_0: \lambda = 0$ versus $H_1: \lambda > 0$
which is equivalent to testing
$H_0: p_b = q$ versus $H_1: p_b \neq q$.
Hence
we are in a two-sample testing scenario 
\citep{kim2016classification, kim2019global}.

Again, we want to leverage the fact that
classifiers have been found to give accurate
estimates of density ratios 
\citep[Ch.14]{hastie2009elements} \citep{goodfellow2014generative, cranmer2015approximating}.
One strategy is to use a classifier like before 
to obtain a likelihood ratio test statistic, 
but this time we estimate the density ratio
$\psi^\dagger = q/p_b$. 
To do this, we train a classifier to differentiate between
the experimental ($\mathcal{W}$) and background events ($\mathcal{X}$)
instead of the signal ($\mathcal{Y}$)  and background events  ($\mathcal{X}$) . 
As mentioned in the introduction, 
this strategy is similar to the one taken by 
\cite{d2021learning} and \cite{d2019learning}, 
who use a neural network to estimate 
the likelihood ratio test statistic.

A second strategy is to use the 
area under the curve statistic (AUC) \citep{hanley1982meaning}
or the misclassification error/misclassification rate (MCE) 
to evaluate the performance of the classifier.
The intuition behind the second strategy is that
in the absence of signal under the null,
a classifier should not be able to differentiate 
between the background and the experimental data.
We discuss both the strategies below.

As before,
let $h$ denote a classifier 
in the combined (background and experimental) sample,
then
\begin{equation}
\psi^\dagger(z) =
\left(\frac{1-\pi}{\pi}\right)
\left(\frac{h(z)}{1-h(z)}\right)
\end{equation}
where now
$\pi = n/(n+m_b)$.
This leads to our estimate of $\psi^\dagger$
\begin{equation}
\hat{\psi^\dagger}(z) =
\left(\frac{1-\pi}{\pi}\right)
\left(\frac{\hat h(z)}{1- \hat h(z)}\right)
\label{eqn::gamma1}
\end{equation}
where $\hat h(z)$
is the trained classifier.

The likelihood ratio as given in (\ref{eqn::LR}) can be rewritten as
\begin{equation}
\frac{\mathcal{L}(\lambda)}{\mathcal{L}(0)} = \prod_i \psi^{\dagger} (W_i).
\end{equation}
Hence, the LRT statistic is given by
\begin{equation}
T = 2 \sum_i \log  \hat{\psi^{\dagger}} (W_i).
\label{eqn::semisuperLRT}
\end{equation}
Similar to the arguments presented in the model-dependent case,
since we are estimating $\psi^{\dagger}$ using $ \hat{\psi^{\dagger}}$,
usual asymptotics do not hold for the LRT statistic.
Instead we propose a conditional asymptotic test. 
We split the background and the experimental data 
into training and test data and estimate $\hat h$
on the training data. We then use the 
test experimental data to calculate the test statistic $T$
and use the test background data to approximate the null distribution. 
Unlike the chi-squared null distribution considered by \cite{d2021learning} and \cite{d2019learning}, we instead use an asymptotic Normal distribution 
to approximate the null distribution. 
In this paper, the split of the data into training and test data sets is performed 
roughly equally. The reason for this is that if the size of the training data is small, we might not have sufficient signal samples in the training data for the classifier to recognize them, and if the size of the test data is small, there might not be enough signal samples in the test set for the test to give a significant result. 
We additionally propose 
nonparametric bootstrap and permutation methods
to approximate the null distribution as well.

For the second strategy that uses the performance of the classifier ($\hat h$) 
in separating the background data from the experimental data,
we consider two statistics - 
AUC (Area Under the Curve) statistic and 
MCE (Misclassification Error) statistic. 

A conventional summary of the ROC curve is
the Area Under the Curve, or AUC. 
The ROC curve \citep{metz1978basic, hanley1989receiver} 
demonstrates the performance of a classifier
by plotting the true positive rate (TPR) versus the false positive rate (FPR) 
at various threshold settings of the classifier output. 
For example, for our classifier $\hat h(z)$, 
given a threshold parameter $t$, 
an instance $Z$ is classified as experimental (``positive") 
if $\hat h(Z) > t$ and 
background (``negative") otherwise. 
Now $Z \sim q$ if $Z$ actually belongs to the experimental class 
and $Z \sim p_b$ if $Z$ actually belongs to the background class. 
Therefore, the TPR is given by 
$\mbox{TPR}(t) = \P_{q} \left(\hat h(Z) > t \right)$ 
and the FPR is given by 
$\mbox{FPR}(t) = \P_{p_b} \left(\hat h(Z) > t \right)$,
where $\P_{q}$ is the probability when $Z \sim q$ and
$\P_{p_b}$ is the probability when $Z \sim p_b$. 
Since, the ROC curve plots $\mbox{TPR}(t)$ (y-axis) versus $\mbox{FPR}(t)$ (x-axis) with varying $t$, 
the AUC $\theta$ is given by:
\begin{equation}
\theta = \int_{0}^{1} \mbox{TPR}(\mbox{FPR}^{-1}(x)) \ dx = \P\left( \hat h(W) > \hat h(X)\right),
\end{equation}
by standard derivation. 
So, the AUC can also be interpreted as 
the probability that a classifier 
will rank a randomly chosen positive instance 
higher than a randomly chosen negative one. 
Hence we can estimate it using
\begin{equation}
\label{eqn::AUC}
\hat\theta = \frac{1}{m_b \ n} \ \sum_i \sum_j \ \mathbb{I}\left\lbrace \hat h(W_j) > \hat h(X_i)\right\rbrace.
\end{equation}

Under the null $H_0: \lambda = 0$, 
we have that
$q = p_b$, i.e., $X$ and $W$ have the same distribution. 
So under $H_0$, $\theta = 0.5$ and an AUC that is significantly greater
than $0.5$ provides evidence of $q \neq p_b$. In other words,
testing $H_0: \lambda = 0$ versus $H_1: \lambda > 0$ is equivalent to
testing $H_0: \theta = 0.5$ versus $H_1: \theta > 0.5$.

We can also use the MCE (misclassification error/classification error rate)
to measure the performance of the classifier \citep{kim2016classification}, 
where we define the MCE as:
\begin{equation}
M = 0.5 \ \P(\hat h(X) > \pi) +  0.5 \ \P(\hat h(W) < \pi).
\end{equation}
Note that this is the average of the false positive rate (first term) 
and the false negative rate (second term) for threshold $\pi = n/(n + m_b)$.
This can be estimated using
\begin{equation}
\label{eqn::MCE}
\hat M =  \frac{1}{2} \left[ \frac{1}{m_b } \ \sum_i  \ \mathbb{I}\left\lbrace \hat h(X_i) > \pi \right\rbrace +
 \frac{1}{n} \  \sum_j  \ \mathbb{I}\left\lbrace \hat h(W_j) < \pi \right\rbrace \right].
\end{equation}
Under the null, $H_0: \lambda = 0$ and as a result, $q = p_b$, i.e., $X_i \overset{d}{=} W_j$. 
Hence,  $M = 0.5$ under the null.
The classifier will have a true accuracy significantly above
half (and hence $M$ below half), only if $q \neq p$. 
Hence we can use $\hat M$ as a test statistic
and test $H_0: M = 0.5$ versus $H_1: M < 0.5$.

As with the LRT statistic, the tests using the AUC and the MCE statistics 
can also be performed using asymptotic, bootstrap and permutation methods. 
The asymptotic AUC method, 
unlike the asymptotic LRT and  MCE methods,
does not use a conditional test.
For the AUC we derive the asymptotic distribution of the statistic
using results presented in \cite{newcombe2006confidence}.
These methods also use data splitting, 
where we use the training data to fit the classifier ($\hat h$)
and use the test data to perform the test.
We detail the nonparametric bootstrap and one of the permutation methods in 
Method~\ref{mthd::fastbootperm} below.  We provide algorithms for the other methods, including the asymptotic methods 
using the three statistics LRT, AUC and MCE, and 
the slower in-sample permutation method in the Supplementary Material \citep{supplementary}.

\begin{Remark}
\begin{enumerate}
\item All the model-independent methods presented in this section, 
except the in-sample permutation method, 
use data splitting of 
the background and the experimental data
into training and test data,
to approximate the null distribution and to 
perform the test. 
This has the benefit that the classifier 
can be kept fixed when estimating the null distribution.
Here splitting a sample means 
randomly splitting the sample into two disjoint subsamples.
The training data is used to train the classifier $\hat h$
and the test data is used to perform the signal detection test.
The in-sample permutation method, 
on the other hand, 
re-trains the classifier on 
permuted background and experimental data 
when approximating the null. 
This has the advantage of using all the data to train the classifier, but the computing time is much longer due to the classifier re-training required when obtaining the null distribution.
\item It is also important to note that here we used the random forest (RF) classifier to predict posterior probabilities. However, the ``vote tallies'' produced by RF classifiers are not posterior probabilities from a generative model, and very often are uncalibrated when interpreted as such. To calibrate the random forest classifier outputs, one can use methods to improve the calibration by using Platt scaling or other more sophisticated methods \citep{mizil2005, henrik2008} on the random forest outputs before performing the tests. We performed several experiments for the model-independent asymptotic tests using Platt scaling on the random forest outputs for calibration and the results were similar to the uncalibrated random forests used here. So we leave the decision of whether to further calibrate the random forest outputs before using the tests to the discretion of the user.
\end{enumerate}
\end{Remark}

\begin{mdframed}
\begin{method}
\label{mthd::fastbootperm}
Bootstrap and Permutation Methods --- faster than the in-sample permutation method, but slower than the asymptotic methods.
\begin{enumerate}
\item Split background data 
${\cal X} = \{X_1,\ldots, X_{m_b}\}$ into
${\cal X}_1$ and ${\cal X}_2$ of sizes $m_1$ and $m_2$ respectively.
\item Split experimental data 
${\cal W} = \{ W_1,\ldots, W_n \}$ into
${\cal W}_1$ and ${\cal W}_2$ of sizes $n_1$ and $n_2$ respectively, 
with $n_2 = m_2$
\item Train the classifier $\hat h$ on ${\cal X}_1$ and ${\cal W}_1$.
\item Evaluate the LRT statistic $T$ 
on ${\cal W}_2$ using (\ref{eqn::gamma1}) and (\ref{eqn::semisuperLRT}) as
\begin{equation}
\tilde{T} = \frac{T}{2 n_2} = \log \left(\frac{1-\pi}{\pi}\right) + \frac{1}{n_2}
\sum_{W_i \in {\cal W}_2} \log \left(\frac{\hat h(W_i)}{ 1- \hat h(W_i)}\right)
\label{eqn::semisuperLambdaasym}
\end{equation}
where $\pi = n_1/(m_1+n_1)$. Similarly evaluate the AUC statistic $\hat{\theta}$, as defined in Equation~(\ref{eqn::AUC}), and the MCE statistic $\hat{M}$, as defined in Equation~(\ref{eqn::MCE}), 
on ${\cal X}_2$ and ${\cal W}_2$ as 
\begin{equation}
\label{eqn::AUCasym}
\hat{\theta} = \frac{1}{m_2 n_2} \sum_{X_i \in {\cal X}_2} \sum_{W_j \in {\cal W}_2}  \mathbb{I}\left\lbrace \hat h(W_j) > \hat h(X_i)\right\rbrace
\end{equation}
\begin{equation}
\label{eqn::MCEasym}
\hat{M} = \frac{1}{2} \left[\frac{1}{m_2} \ \sum_{X_i \in {\cal X}_2} \ \mathbb{I}\left\lbrace \hat h(X_i) > \pi \right\rbrace + \frac{1}{n_2} \  \sum_{W_j \in {\cal W}_2}  \mathbb{I}\left\lbrace \hat h(W_j) < \pi \right\rbrace \right]
\end{equation}
respectively.
\item  Estimate the null distribution of $\tilde{T}$, $\hat{\theta}$ and $\hat{M}$, 
by repeatedly drawing $m_2 + n_2$ random observations
 from ${\cal X}_2 \cup {\cal W}_2$
(with replacement for bootstrap and without replacement for permutation)
and randomly labelling $m_2$ of them as $X$'s and $n_2$ of them as $W$'s
before computing $\tilde{T}$, $\hat{\theta}$ and $\hat{M}$ on them. (Note that under the null, $q = p_b$ implying that the $X$'s and the $W$'s have the same distribution.)
\item Calculate the p-values based on the estimated null distributions.
\end{enumerate}
\end{method}
\end{mdframed}

\section{Estimating $\lambda$ in the Model Independent Scenario}
\label{sec::lambdaEst}

Here we discuss the problem of estimating the signal strength $\lambda$ 
in the semi-supervised setting.
As we have seen in Section~\ref{sec::semisupervised},
the ratio of the densities $\psi^\dagger = q/p_b$ can be written as:
\[ \psi^\dagger(z) =
\left(\frac{1-\pi}{\pi}\right)
\left(\frac{h(z)}{1-h(z)}\right), \]
where
$\pi = n/(n+m_b)$ and 
$h$ is the classifier differentiating the experimental data from the background data.
Since
\begin{equation}
\psi^\dagger(z) = \frac{q(z)}{p_b(z)} = \frac{(1 - \lambda) p_b(z) + \lambda p_s(z)}{p_b(z)} = 1-\lambda + \lambda \psi(z),
\label{eqn::gamma2}
\end{equation}
we see that, for any $z$ such that
$\psi(z)\neq 1$ we have
\begin{equation}
\lambda =
\frac{1-\left(\frac{1-\pi}{\pi}\right)
\left(\frac{h(z)}{1-h(z)}\right)}{1-\psi(z)}.
\end{equation}
Hence,
if we can find any $z$ for which $p_s(z)=0$ (no signal) then
we can estimate $\lambda$ by
\begin{equation}
\hat\lambda =
1-\left(\frac{1-\pi}{\pi}\right)
\left(\frac{\hat h(z)}{1-\hat h(z)}\right),
\end{equation}
where $\hat{h}(z)$ is the trained semi-supervised classifier.
The problem is that the search for such a $z$ may not be obvious.

Instead we take a different approach.
To ensure identifiability, we assume $\inf_{z} p_s(z)/p_b(z) = 0$. 
We believe this is true in most high-energy physics problems.
To simplify the discussion, we also assume $p_b, q > 0$ everywhere.

Next we show that the problem of estimating $\lambda$
can be transformed into a problem of estimating $g_q(1)$,
where $g_q$ is the density of a univariate random variable 
supported on $[0, 1]$. 
We define for any 
$t \geq 0$, $C_t = \{ z \in \mathbb{R}^d : p_s(z) \geq t p_b(z)\}$. 
Then for any $z \in \mathbb{R}^d$, 
we define the Neyman-Pearson Quantile Transform of $z$ as:
\begin{equation}
\rho(z) =  \P_{X \sim p_b} \left(\psi^\dagger(X) \geq \psi^\dagger(z)  \right)  =  \P_{X \sim p_b} \left(h(X) \geq h(z)  \right).
\end{equation}

Let $g_0$, $g_1$ and $g_q$ be the density functions of $\rho(Z)$ when $Z \sim p_b$, $Z \sim p_s$ and $Z \sim q$ respectively. Then we can show, as stated in Theorem~\ref{thm::Lambda} below, that $g_q(1) = 1 - \lambda$ and therefore $\lambda = 1 - g_q(1)$.
So, we can estimate $\lambda$ using
\begin{equation}
\hat\lambda = 1 - \hat{g_q}(1), 
\end{equation}
where $\hat{g_q}$ is a density estimate based on the $\hat\rho(W_i)$'s defined as
\begin{equation}
\hat\rho(W_i) = \frac{1}{m_b} \sum_{j } \mathbb{I}\left\lbrace \hat{h}(X_j) \geq \hat{h}(W_i)\right\rbrace
\label{eqn::rhohat}
\end{equation}

\begin{theorem}
\label{thm::Lambda}
Assuming $\rho(Z)$ has continuous density under either $Z \sim p_b$ or  $Z \sim p_s$, then the following hold.
\begin{enumerate}
\item $g_0(u) = 1$ for all $u \in [0, 1]$. That is, $\rho(Z) \sim \mathrm{Unif}(0, 1)$ if $Z \sim p_b$.
\item $g_1(u) = t_u$ where $t_u$ satisfies $P_{X \sim p_b}(X \in C_{t_u}) = u$. In particular, $g_1(1) = 0$.
\item $g_q(1) = 1 - \lambda$.
\end{enumerate}
\end{theorem}

We detail the proof of the Theorem in Section 1 of the Supplementary Material \citep{supplementary}.

Now the problem of estimating $\lambda$ reduces to estimating a monotone density at a boundary point. We can estimate the density $g_q(1)$ based on the $\hat{\rho}(W_i)$'s using a simple histogram based estimator:
\begin{equation}
\label{eqn::densityestimate}
\hat{g_q}(1) = \frac{1}{nb} \sum_i \mathbb{I}\left\lbrace \hat{\rho}(W_i) \in (1 - b, 1] \right\rbrace,
\end{equation}
where $b$ is the bin-width of the histogram estimator. But, since the density is a monotonically decreasing function, the density estimates at points close to 1 could also be indicative of the estimate at 1. 

We therefore propose using a Poisson regression on bins close to 1, in order to estimate the density at 1. We fit a Poisson regression $\hat{f}(t) = \exp(\beta_0 + \beta_1 t),$ with $\beta_1 \leq 0$ to the histogram estimates
\begin{equation}
\label{eqn::histogramest}
H_t =  \sum_i \mathbb{I}\left\lbrace \hat{\rho}(W_i) \in (t - b, t] \right\rbrace, T \leq t - b \leq t  \leq 1,
\end{equation}
where $b$ is the bin-width of the histogram estimator and $T$ determines the neighborhood of 1 that is used to estimate the density at 1. 
Then the estimated density at 1 is given by:
\begin{equation}
\label{eqn::densityestimateGLM}
\hat{g_q}(1) = \hat{f}(1),
\end{equation}
the estimate given by the Poisson regression at 1.

\begin{Remark}

\begin{enumerate}
\item Note that the bins $(t-b, t]$ for $T \leq t - b \leq t  \leq 1$ 
form a partition of $[T, 1]$ and we regress on the bin end points 
for the Poisson regression model. 
\item We constrain $\beta_1 \leq 0$, since we know that $g_q$ 
is a monotonically decreasing function. 
In practice, we implement this condition by 
setting $\widehat{\beta_1} = 0$, 
when the Poisson regression models estimates $\widehat{\beta_1} > 0$ 
and fit $\hat{f}(t) = \exp(\beta_0)$. 
\item In practice, we additionally perform data-splitting in order to get
out-of-sample estimates of $\lambda$. 
It's important to consider out-of-sample estimates 
since the Poisson regression model is conditioned on the trained classifier 
and computing the signal strength estimates on the same data that is used 
to train the classifier could give biased estimators. 
\item Instead of fitting a Poisson regression model on the histogram estimates $H_t$, we additionally tried using just the mean of the density estimates given by the histogram above a threshold $T$ to estimate $\hat{f}(1)$. We also tried using a simple linear regression on $\log(H_t)$. These methods experimentally gave poorer estimates compared to the Poisson model. 
\end{enumerate}
\end{Remark}

We compute the out-of-sample estimate of $\lambda$ as mentioned in Method~\ref{mthd::est} below. We can additionally use nonparametric bootstrap 
to understand the stability of the signal strength estimates to 
perturbations in the data. 
We detail the bootstrap process below in Method~\ref{mthd::estboot}.  
The method provides standard error estimates as well as 
bootstrapped confidence intervals that 
characterize the stability of the estimated signal strength $\lambda$.

\begin{mdframed}
\begin{method}
\label{mthd::est}
Estimating the Signal Strength $\lambda$.
\begin{enumerate}
\item Split background data 
${\cal X} = \{X_1,\ldots, X_{m_b}\}$ into
${\cal X}_1$ and ${\cal X}_2$ of sizes $m_1$ and $m_2$ respectively.
\item Split experimental data 
${\cal W} = \{ W_1,\ldots, W_n \}$ into
${\cal W}_1$ and ${\cal W}_2$ of sizes $n_1$ and $n_2 = m_2$ respectively.
\item 
Train the classifier $\hat h$ on ${\cal X}_1$ and ${\cal W}_1$.
\item Compute $\hat{\rho}(W_i)$ for all $W_i \in {\cal W}_2$ as defined in (\ref{eqn::rhohat}) based on $\hat h$ and ${\cal X}_2$ as
\begin{equation}
\hat\rho(W_i) = \frac{1}{m_2} \sum_{X_j \in {\cal X}_2} \mathbb{I}\left\lbrace \hat{h}(X_j) \geq \hat{h}(W_i)\right\rbrace, \quad W_i \in {\cal W}_2.
\end{equation}
\item Get histogram estimates $H_t$ of $\hat\rho(W_i)$'s for bins larger than $T$ with bin-width $b$ using Equation~(\ref{eqn::histogramest}). 
\item Use Poisson regression to estimate the density at 1 as $\hat{g_q}(1)$. Then $\hat{\lambda} = 1 - \hat{g_q}(1)$.
\end{enumerate}
\end{method}
\end{mdframed}

\begin{mdframed}
\begin{method}
\label{mthd::estboot}
Bootstrapped Uncertainty Intervals for $\hat{\lambda}$.
\begin{enumerate}
\item  Repeatedly draw with replacement $m_b$ background samples from $\mathcal{X}$ and $n$ experimental samples from $\mathcal{W}$ and split them into ${\cal X}_1^*$, ${\cal X}_2^*$, ${\cal W}_1^*$ and ${\cal W}_2^*$ of sizes $m_1, m_2, n_1$ and $n_2$ at random ensuring no overlap between the training and test data sets. That is, ${\cal X}_1^* \cap {\cal X}_2^* = \phi$ and ${\cal W}_1^* \cap {\cal W}_2^* = \phi$. $m_1, n_1, m_2$ and $n_2$ are same as ones used in Method~\ref{mthd::est} above.
\item Find $\hat{\lambda}^*$ in each case using steps 3-6 in Method~\ref{mthd::est}. Note that the classifier is re-trained on ${\cal X}_1^* \cup {\cal W}_1^*$ for every random sample.
\item We can then use the empirical standard deviation or quantiles of the $\hat{\lambda}^*$'s to create bootstrap confidence intervals that will give estimates of the stability of the estimator $\hat{\lambda}$ to perturbations in the data.
\end{enumerate}
\end{method}
\end{mdframed}

\begin{Remark}
Since the data are split to consider an out-of-sample estimate of $\lambda$,
it is also necessary in the bootstrapped samples 
for the training set to be disjoint of the test set, 
to avoid problems caused by considering an in-sample estimate.
\end{Remark}

\section{Interpreting the Classifier}
\label{sec::active}

The signal detection test relies on
the classifier; here we use Random Forest.
So, to understand, characterize and interpret 
the new physics signal detected by the test,
it is useful and necessary to find a way to interpret the fitted classifier. 
In this section, 
we propose two methods to 
help interpret the random forest classifier.

The methods are based on understanding 
how the gradient of the classifier surface ($\hat{h}(z)$) 
varies. 
The underlying idea is that 
directions in which the surface is sloped 
or changes a lot 
contain useful information for the classification, 
while directions in which the surface is flat do not.
For the first method, 
we look at the density of the gradients
of the classifier marginally along each of the variables.
For the second method, we find 
multivariate dependencies that jointly affect the gradient of the classifier,
via finding the active subspace \citep{constantine2015active} 
of the classifier surface. 
The active subspace is found by looking at 
the leading eigenvectors of the standardized gradients 
of the classifier surface by performing PCA on the standardized gradients.

\begin{figure}[h]
\begin{subfigure}{.98\textwidth}
  \centering
   \includegraphics[width=\linewidth]{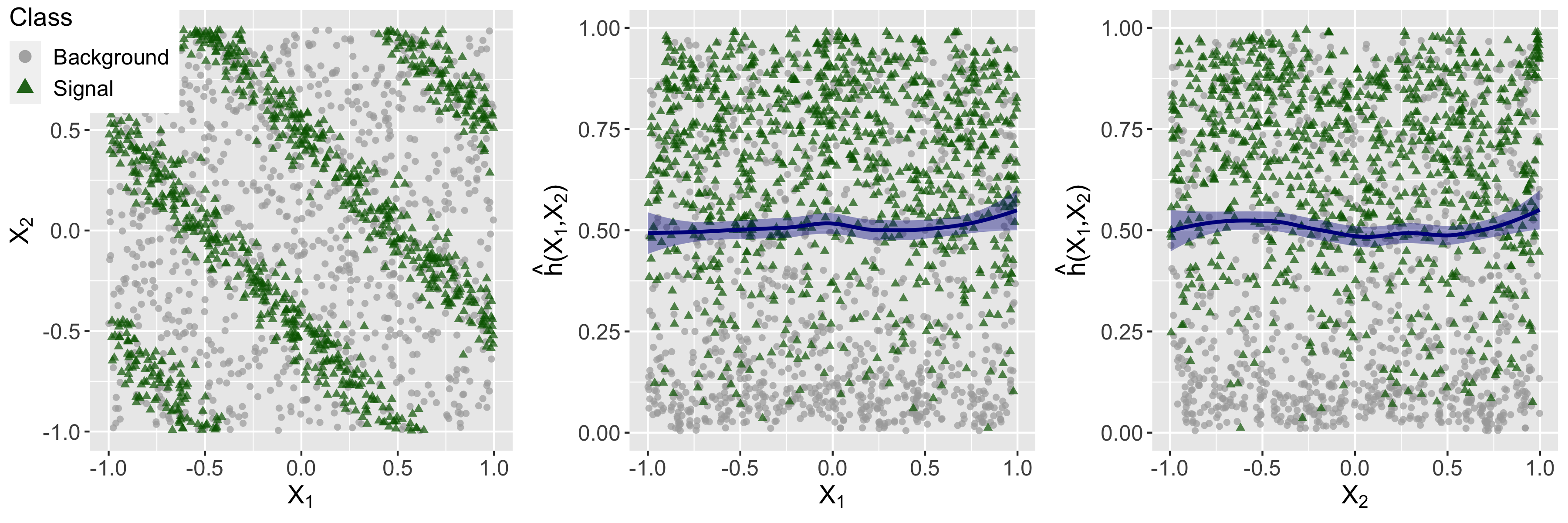}
  \caption{$X_1$ versus $X_2$, $\hat{h}(X_1, X_2)$ versus $X_1$ and $\hat{h}(X_1, X_2)$ versus $X_2$}
  \label{fig:ploteg}
\end{subfigure}
\begin{subfigure}{.44\textwidth}
  \centering
   \includegraphics[width=\linewidth]{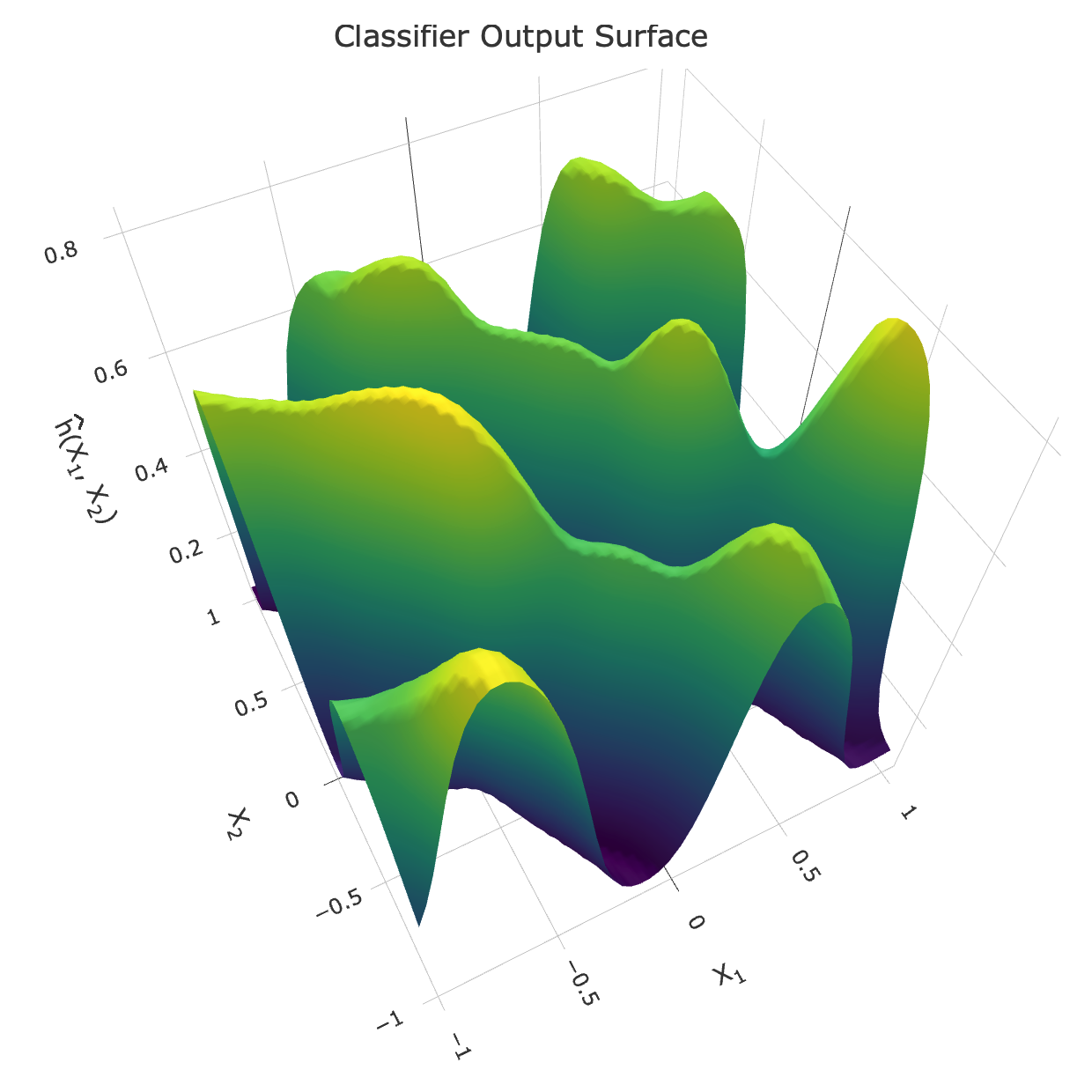}
  \caption{Smoothed Classifier Surface}
  \label{fig:surfaceeg}
\end{subfigure}
\begin{subfigure}{.44\textwidth}
  \centering
   \includegraphics[width=\linewidth]{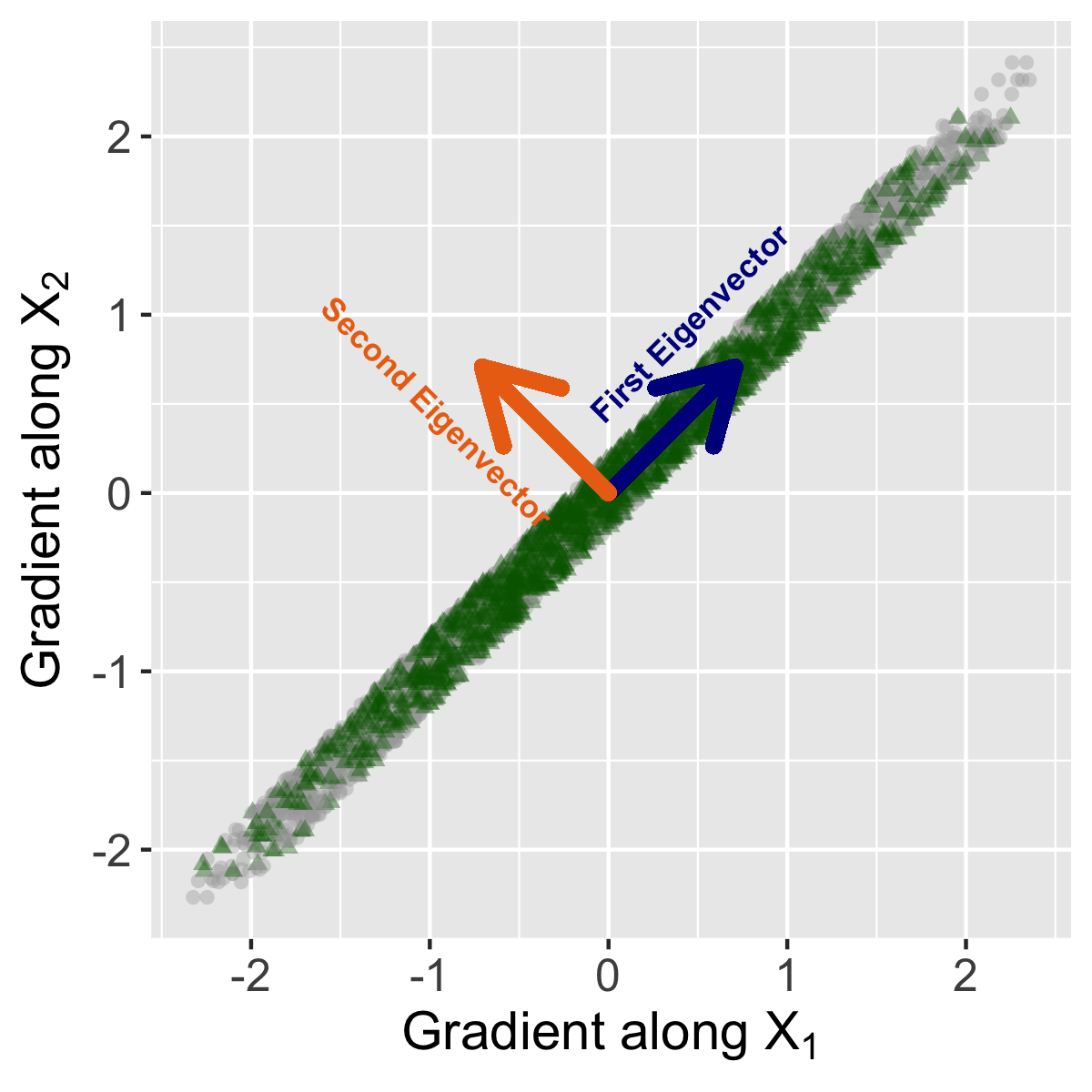}
  \caption{PCA of the Standardized Gradients}
  \label{fig:PCAeg}
\end{subfigure}
 \caption{We demonstrate the active subspace method on a two-dimensional example as shown in (a), where grey circles denote the uniformly distributed background and the green triangles denote the signal. The signal lies around lines parallel to the line $X_1 + X_2 = 0$.  Additionally, (a) also shows the output of the classifier that separates the signal from the background, $\hat{h}(X_1, X_2)$, as a function of $X_1$ and $X_2$ individually. Neither of these variables marginally has separation power for the signal. (b) shows the smoothed classifier surface as a function of both $X_1$ and $X_2$. The active subspace method then performs PCA on the standardized gradients of the smoothed classifier output in (b). (c) shows a scatter plot of the standardized gradients of the smoothed classifier output as well as the two eigenvectors. We see that, as expected, the first eigenvector picks the direction in which the  classifier output varies the most, namely $X_1 - X_2 = 0$. We consider a two-dimensional example here for illustration purposes only. The real data can be of much higher dimensionality.}
 \label{fig::exampleactive}
\end{figure}

Figure~\ref{fig::exampleactive} demonstrates the 
active subspace method on a two-dimensional example for simplicity, 
where the data $(X_1, X_2) \in [-1, 1] \times [-1, 1]$ and 
the signal lies around lines parallel to the line $X_1 + X_2 = 0$. 
We then train a classifier $\hat{h}(X_1, X_2)$ to separate the signal from the background. 
We notice that the classifier output does not appear to have any 
relationship with  either $X_1$ or $X_2$ marginally, as shown in Figure~\ref{fig:ploteg}. 
But the smoothed classifier surface detects the signal around 
lines parallel to the line $X_1 + X_2 = 0$, as can be seen by the ridges 
along those lines in Figure~\ref{fig:surfaceeg}. 
So, looking at the standardized gradients of the classifier surface  
gives us information about the direction in which the surface changes. 
As seen in Figure~\ref{fig:PCAeg}, the first eigenvector of the standardized gradients 
reflects the direction in which the classifier surface changes the most, 
which is along the $X_1 - X_2 = 0$ line. 
Identifying this subspace helps us understand the directions in the feature space that separate the signal from the background.  
Note that we consider a two-dimensional example here for illustration purposes only. The real data can be of much higher dimensionality.

In what follows, instead of directly looking at the classifier $\hat{h}(z)$, 
we look at $\text{logit}(\hat{h}(z)) = \log\left(\hat{h}(z)/ (1 - \hat{h}(z))\right)$. 
In practice for computational purposes, we add a small noise (e.g. $10^{-10}$) to $\hat{h}(z)$ 
when $\hat{h}(z) \in \{0, 1\}$ to ensure that $\text{logit}(\hat{h}(z))$ stays finite. 
We notice that taking the logit transformation 
provides more stable estimates of the gradients of the surface. 
Henceforth, we will instead look at the gradients of $H(z) := \text{logit}(\hat{h}(z))$.

To find the gradient $\nabla_{z} H(z)$,
we fit a local linear smoother to the logit of the random forest output. 
That is, we fit the logit of the classifier outputs 
$H(Z_i) = \text{logit}(\hat{h}(Z_i))$ locally around $z^*$ using
\begin{equation}
\label{eqn::linearsmoother}
\hat{H}(z) = \alpha(z^*) +  \beta( z^* )^T(z - z^*) + o\left(|| z - z^*||_2^2\right).
\end{equation}

Then, $\hat\beta(Z_i)$ provides estimates of the gradient 
of the logit classifier output on the data.
We furthermore replace the gradient estimates $\hat\beta_j(Z_i)$
by their standardized versions $\hat\beta_j(Z_i)/sd(\hat\beta_j(Z_i))$
at every data point $Z_i$, to get more stabilized values.
So, henceforth $\hat\beta(Z_i)$ will be used 
to indicate the standardized gradient estimates
for notational simplicity.

\begin{Remark}
The data used for estimating the gradient 
can either be the background data or 
the experimental data or a combination of both.
We use a combination of both since 
experimentally we see that 
using the combination captures the 
classifier surface better.
\end{Remark}

We can then look at:
\begin{itemize}
\item[(i)] The density of $\hat\beta_j(Z)$, 
the estimated standardized gradient of $H(Z)$ along the $j^{th}$ variable.
\item[(ii)] The active subspace 
found using the estimated standardized gradients $\hat\beta(Z)$ 
as detailed in Section~\ref{sec::active} below.
\end{itemize}

The density of $\hat\beta_j(Z)$ explains how the classifier behaves 
marginally along each variable, 
whereas the active subspace identifies 
relationships between multiple variables that 
cause the most change in the classifier. 
We now describe the active subspace method in detail below.

\subsection{Active Subspace of the Classifier}
\label{subsec::activesubspace}
In this section we find the active subspace of the classifier $\hat h$,  
which is assumed to be fixed and all the expectations 
in this section are with respect to the inputs of the fixed classifier.
Let us consider the mean and the covariance matrix of $\nabla_z H(z)$. 
We define
\begin{equation}
C = \E_f\left[\left(\nabla_{ z} H - \E\left[\nabla_{ z} H \right]\right) \left(\nabla_{ z} H -  \E\left[\nabla_{ z} H \right]\right)^T\right],
\end{equation}
where $\E_f$ is the expectation with respect to the density 
$f = c p_b + (1 - c) q$, where
$c = m_2/(m_2 + n_2)$ gives the proportion of background data 
in the combined sample of background data $\mathcal{X}_2$ of size $m_2$ 
and experimental data $\mathcal{W}_2$ of size $n_2$ and
$q = (1 - \lambda) p_b + \lambda p_s$. 
We consider expectations with respect to $f$ since 
as explained in the remark above, we use a combination 
of the experimental data and the background data to 
find the active subspace of the classifier.

The mean $\E_f\left[\nabla_{ z} H \right]$ 
gives the expected slope of the surface of $H(z)$.
For example, a positive mean gradient along the $j^{th}$ variable
indicates an increasing trend in the classifier output with that variable, 
meaning that increasing that variable increases the probability of an experimental data point and decreases the probability of a background data point. 
The covariance matrix $C$, on the other hand, encodes the 
variable relationships that cause variability in the classifier surface.

Consider the real eigenvalue decomposition of $C$,
\begin{equation}
\label{eqn::CEigen}
C = M \Lambda M^T, \ \ \ \Lambda = \text{diag}(\lambda_1, \ldots, \lambda_d), \ \ \lambda_1 \geq \ldots \geq \lambda_d \geq 0,
\end{equation}
where $M$ has columns $\{{M}_{\cdot 1}, \ldots, {M}_{\cdot d}\}$, the
normalized eigenvectors of $C$.  
Then the vector ${M}_{\cdot 1}$ corresponding to
$\lambda_1$, gives the association between the variables 
(i.e., the direction in the input space) 
that best captures the changes in $H(z)$ around the expected slope, 
followed by
${M}_{\cdot 2}$ and so on. 
Therefore the eigenvectors corresponding to
the leading eigenvalues $\lambda_1, \lambda_2, \ldots$, 
give us an idea about the directions along which 
the classifier output changes the most. 
These are directions that contain meaningful information for separating the experimental data from the background data and therefore enable us to characterize how the experimental data differs from the background data. 
Towards this end, we propose Method~\ref{mthd::active}  
that uses the gradient estimates $\hat\beta(Z)$ derived from
a local linear smoother.

\begin{mdframed}
\label{mthd::active}
\begin{method} Active Subspace for the classifier $\hat h$.
\begin{enumerate}
\item Split background data into
${\cal X}_1$ and ${\cal X}_2$ of sizes $m_1$ and $m_2$ respectively.
Split experimental data into
${\cal W}_1$ and ${\cal W}_2$ of sizes $n_1$ and $n_2 = m_2$ respectively.
Train the classifier $\hat h$ on ${\cal X}_1$ and ${\cal W}_1$.
\item Fit a local linear smoother that estimates $H(Z_i) = \text{logit}\left(\hat h(Z_i)\right)$ for $Z_i \in {\cal X}_2 \cup {\cal W}_2$ using the model given in (\ref{eqn::linearsmoother}).
\item Consider the coefficients of the local linear smoother
$\hat\beta(Z_i)$ for $Z_i \in {\cal X}_2 \cup {\cal W}_2$. 
They provide an estimate of the gradient $\nabla_{ z} H$ at $Z_i$, i.e., 
$\widehat{\nabla_{ z} H(Z_i)} = \hat\beta(Z_i)$ for $Z_i \in {\cal X}_2 \cup {\cal W}_2$.
\item Standardize the estimated gradients $\hat\beta_j(Z_i)$ to  
$\hat\beta_j(Z_i)/\widehat{sd(\hat\beta_j(Z_i))}$ 
by using the estimated $\widehat{sd(\hat\beta_j(Z_i))}$ 
given by the local linear smoother.
\item Then instead of $C$ we can use the estimate
\begin{equation}
\widehat{C} = \frac{1}{N} \sum_{Z_j \in {\cal X}_2 \cup {\cal W}_2} \left( \hat\beta(Z_j) - \overline{\hat\beta(Z)}\right) \left(\hat\beta(Z_j) - \overline{\hat\beta(Z)} \right)^T,
\label{eqn::CHat}
\end{equation}
where $\overline{\hat\beta(Z)} =  \sum_{Z_j \in {\cal X}_2 \cup {\cal W}_2} \hat\beta(Z_j) / N$ and $N = m_2 + n_2$.

\item Find the eigenvalue decomposition of $\widehat{C}$ as
\[ \widehat{C} = \widehat{M} \widehat{\Lambda}  \widehat{M}^T,\]
which gives the estimates $ \widehat{M}$ and $\widehat{\Lambda}$ of $M$ and $\Lambda$ as defined in (\ref{eqn::CEigen}) respectively. We find the eigenvectors by performing PCA on the standardized gradients.

\item Then $\overline{\hat\beta(Z)}$ and the estimated eigenvectors $\hat{M}_{\cdot 1}, \hat{M}_{\cdot 2}, \ldots$ given by the columns of $ \widehat{M}$, 
best capture the slope and variations in $H(\cdot)$ and hence the classifier surface, respectively.
\end{enumerate}
\label{mthd::active}
\end{method}
\end{mdframed}

We can additionally construct bootstrapped uncertainty intervals by repeatedly drawing with replacement $m_b$ background samples from ${\cal X}$ and $n$ experimental samples from  ${\cal W}$ and split them into ${\cal X}_1^*$, ${\cal X}_2^*$, ${\cal W}_1^*$ and ${\cal W}_2^*$ of sizes $m_1, m_2, n_1$ and $n_2$ at random ensuring no overlap between the training and test data sets. That is, ${\cal X}_1^* \cap {\cal X}_2^* = \phi$ and ${\cal W}_1^* \cap {\cal W}_2^* = \phi$. In each case, we first re-train the classifier on the new training data ${\cal X}_1^* \cup {\cal W}_1^*$ and then find $\overline{\hat\beta(Z)}$, and the estimated eigen values $\hat{M}_{\cdot 1}, \hat{M}_{\cdot 2}, \ldots$ on the new test data ${\cal X}_2^* \cup {\cal W}_2^*$. Here $m_1, m_2, n_1$ and $n_2$ are the same as in Method~\ref{mthd::active}. We can then use the standard deviation or quantiles of the bootstrapped estimates to create bootstrap confidence intervals that will give estimates of the stability of the estimators, given by the algorithm, to perturbations in the data.

\section{Experiments: Search for the Higgs Boson}
\label{sec::higgs}

We demonstrate the performance of the proposed 
semi-supervised classifier tests 
on the Higgs boson machine learning challenge data set 
available on the CERN Open Data Portal at 
\url{http://opendata.cern.ch/record/328} \citep{atlashiggs}.
The data set consists of simulated data provided by the ATLAS experiment
at CERN's Large Hadron Collider to optimize the search for the Higgs boson. 

Our goal is to demonstrate the performance of the proposed tests
in identifying the presence of the Higgs boson particle 
without assuming an a priori ansatz of the signal and 
demonstrate their applicability to model-independent searches 
of new physics signals in experimental particle physics.


\subsection{Data Description}

The Higgs boson has many different ways through which
it can decay in an experiment and produce other particles. 
This particular challenge, from which our data set originates, 
focuses on the collision events where 
the Higgs boson decays into two tau particles \citep{pmlr-v42-cowa14}. 
The data provided for the challenge consist of collision events labelled as background 
and signal. 
The signal class is comprised of events 
in which a Higgs boson (with a fixed mass of 125 GeV) 
was produced and then decayed
into two taus. 
The events are simulated using the official ATLAS full detector simulator.
The simulator yields simulated events with properties that mimic the statistical
properties of the real events of the signal type as well as several important backgrounds.
For the sake of simplicity, 
background events generated 
from only three different background processes 
were included in the challenge data set.
Our objective is to show that semi-supervised classifier tests
are able to identify the Higgs signal without any prior knowledge. 

The data set has 818,238 observations,
where each observation is a simulated 
proton-proton collision event
 in the detector.
Each of these collision events produces 
clustered showers of hadrons, 
which originate from a quark or a gluon 
produced during the collision.
These showers are called \emph{jets}.
The data contain information on the measured properties of the jets 
as well as the other particles produced during the collision.
There are $d = 35$ features whose individual details can be found on \href{http://opendata.cern.ch/record/328}{CERN's Open Data Portal} or in 
Appendix B of \cite{pmlr-v42-cowa14}. 
Here we give some insight into the most important characteristics of the features.

The features whose names start with {\tt PRI} 
are primitive variables that record the raw quantities 
as measured by the detector. 
The features whose names start with {\tt DER} 
are derived variables which are evaluated as 
functions of the primitive variables. 
Since all collision events do not produce the same number of jets,
the number of jets produced in the collisions, denoted by 
\verb!PRI_jet_num!, ranges from $0 - 3$ 
(events with more than $3$ jets are capped at $3$). 
Note that it is possible for the collisons 
to not produce any jets (\verb!PRI_jet_num!$ = 0$) 
and hence there are structurally absent missing values in the data 
that relate to the jets produced in the collisions.
To avoid these missing values in the data,
we only consider events that have two jets (\verb!PRI_jet_num!$ = 2$).
This results in 165,027 events; 
80,806 background events and 84,221
signal events.
Since the derived quantities are
functions of the primitives, we use just the primitive variables ($d =
16$) for our analysis.  
Since we only use the primitive variables, we drop the pre-fix \verb!PRI! from the variable names and further shorten some of the variable names intuitively for convenience. 
Descriptions of the primitive variables used are provided in Table~\ref{tbl::variables}. 

Among the primitive features, 
five of them provide the azimuth angle $\phi$ 
of the particles generated in the event 
(variables ending with \verb!_phi!).
These features are rotation invariant in the sense that
the event does not change if
all of them are rotated together by the same angle.
Hence to interpret these variables more easily using the active subspace method,
we remove the invariance of the azimuth angle variables by rotating all the $\phi$'s 
so that the azimuth angle of the leading jet at $0$ (\verb!lead_phi!$ = 0$). 
\verb!lead_phi! can then be removed from the analysis, 
leading to a 15-dimensional feature space.

\begin{table}[H]
\caption{Descriptions of the variables used in the analysis of the Higgs boson machine learning challenge data set \citep{pmlr-v42-cowa14}.  We drop the pre-fix \texttt{PRI} from the variable names and further shorten some of the variable names intuitively for convenience.}
\label{tbl::variables}
\begin{tabular}{r p{0.7\linewidth}}
\toprule
Variable & Description \\ \midrule
 \verb!tau_pt! &	The transverse momentum of the hadronic tau. \\[3pt]
\verb!tau_eta!	& The pseudorapidity $\eta$ of the hadronic tau. \\[3pt]
\verb!tau_phi! &	The azimuth angle $\phi$ of the hadronic tau. \\[3pt]
\verb!lep_pt!	&The transverse momentum of the lepton (electron or muon). \\[3pt]
\verb!lep_eta!	& The pseudorapidity $\eta$ of the lepton. \\[3pt]
\verb!lep_phi!	& The azimuth angle $\phi$ of the lepton.\\[3pt]
\verb!met!	& The missing transverse energy. \\[3pt]
\verb!met_phi!	& The azimuth angle $\phi$ of the missing transverse energy.\\[3pt]
\verb!met_sumet! &	The total transverse energy in the detector.\\[3pt]
\verb!lead_pt! &	The transverse momentum of the leading jet, i.e., the jet with the largest transverse momentum (undefined if \verb!PRI_jet_num! $= 0$).\\[3pt]
\verb!lead_eta! &	The pseudorapidity $\eta$ of the leading jet (undefined if \verb!PRI_jet_num! $= 0$).\\[3pt]
\verb!lead_phi! & The azimuth angle $\phi$ of the leading jet (undefined if \verb!PRI_jet_num! $= 0$).\\[3pt]
\verb!sublead_pt! & The transverse momentum of the subleading jet, i.e., the jet with the second largest transverse momentum (undefined if \verb!PRI_jet_num! $\leq 1$).\\[3pt]
\verb!sublead_eta!	 & The pseudorapidity $\eta$ of the subleading jet (undefined if \verb!PRI_jet_num! $\leq 1$).\\[3pt]
\verb!sublead_phi! & The azimuth angle $\phi$ of the subleading jet (undefined if \verb!PRI_jet_num! $\leq 1$).\\[3pt]
\verb!all_pt! & The scalar sum of the transverse momentum of all the jets in the event.\\[3pt]
Weight	& The event weight. \\[3pt]
Label &	The event label (string) (s for signal, b for background). \\
\bottomrule
\end{tabular}
\end{table}

Additionally, we take logarithmic transformations of the variables
that give the transverse momentum of the particles produced (variables ending with \verb!_pt!),
the missing transverse energy (\verb!met!) and 
the total transverse energy in the detector (\verb!met_sumet!).
Exploratory analysis of the data 
as well as details and justifications 
for the transformations considered above, 
can be found in the Supplementary Material \citep{supplementary}.

\emph{Experimental Setting:} 
In all the following experiments on the Higgs boson data set, 
we randomly sample without replacement 
background and signal events from the original data set 
to form
background data with $m_b = $ 40,403  events, 
signal data with $m_s = $ 20,403 events, and
experimental data with $n = $ 40,403 events. 
In this manner we generate 50 replicates 
of each of these data sets for use in our power studies.

Recall that the experimental data is a mixture of 
background and signal data, 
where the proportion of signal data 
is given by $\lambda$, the signal strength.
So, the number of signal events in the experimental data
is decided according to $\lambda$
by randomly generating from Bin$(n, \lambda)$. 
For methods that require data-splitting, 
the data-splitting is done by 
randomly splitting the background data into 
$m_1 = $ 20,403 training and 
$m_2 = $20,000 test background samples and 
by randomly splitting the experimental data into 
$n_1 = $20,403 training and  
$n_2 = $20,000 test experimental samples. 
We construct the splits so the number of signal events 
in the training and test experimental samples
are randomly generated from 
Bin$(n_1, \lambda)$ and Bin$(n_2, \lambda)$,
respectively. 
Additionally, when randomly generating the experimental data,
events are selected in a weighted fashion using the Weight variable
provided in the data set as described in Table~\ref{tbl::variables}.
Note that the classifiers for the model-dependent approaches are trained on 
$m_1 = $ 20,403 background and $m_s = $ 20,403 signal training samples, and 
the classifiers for the model-independent approaches are trained on 
$m_1 = $ 20,403 background and $n_1 = $ 20,403 experimental training samples. 
Hence, all the classifiers are trained on balanced datasets. 
For each of the bootstrap and permutation methods 
we consider 1,000 bootstrap and permutation cycles respectively. 

In the following sections, 
we first explore the power of the classifier tests 
described in Section~\ref{sec::methods}
to detect the presence of the Higgs boson signal in the experimental data. 
We then estimate the signal strength ($\lambda$) 
using the methods introduced in Section~\ref{sec::lambdaEst} 
and then use the active subspace method 
introduced in Section~\ref{sec::active} to characterize the signal. 
All the code used for this section is available at \url{https://github.com/purvashac/MIDetectionClassifierTests}.

\subsection{Anomaly Detection Using the Classifier Tests}
\label{sec::testperformance}

We compare the power of the 
model-dependent supervised methods 
and the model-independent semi-supervised methods introduced in
Section~\ref{sec::methods} in detecting the Higgs boson
signal, by varying the signal strength from $\lambda = 0.15$ to
$\lambda = 0.01$. 
We also check that the tests have the right error
control under the null case ($\lambda = 0$). 
We demonstrate the performance 
of the different methods --
asymptotic, bootstrap,
permutation (using out-of-sample statistics),
and slow permutation (using in-sample statistics),
used along with the different test statistics -- 
Likelihood Ratio  Test statistic (LRT), Score statistic, 
Area Under the Curve (AUC), and Misclassification Error (MCE).

For the model-dependent methods, 
since we have only finitely many samples from the background available,
and not the background generator itself, 
we split the available background data into training and test sets as described above. 
We then use the training set for fitting the classifier and 
the test set for estimating 
the null distribution of the test statistic 
by bootstrapping or permuting as described in Section~\ref{sec::methods}. 
In real life, since the background MC generator is known, 
we should, in many cases, 
be able to generate more training background samples 
for estimating the null distribution of the test statistic.  

\subsubsection{Anomaly Detection when Data has a Correctly Specified Signal}

We first compare the methods when the signal model is correctly specified. 
The tests are run on $50$ random samplings of the data, 
and the percentage of times each of the tests rejects the null that there is no signal, 
is given in Table~\ref{tbl::Power}. 
``Permutation'' indicates the faster permutation method from Section~\ref{sec::semisupervised}, 
that uses out-of sample test statistics for testing,  
and ``Slow Perm'' indicates the slower permutation method from Section~\ref{sec::semisupervised} 
that uses in-sample test statistics for testing 
and re-trains the classifier in every permutation cycle. 
The significance level for all the tests is considered at $\alpha = 0.05$.

\begin{table}[h]
\caption{Power of detecting the signal in the Higgs boson data for each method  in percentages. We consider 50 random samplings of the data and 1,000 bootstrap and permutation cycles. We perform each test at $\alpha = 5\% $ significance level. The last column ($\lambda = 0$) represents the Type I error of the methods.}
\label{tbl::Power}
\begin{tabular}{@{}lcrrrrrrr@{}}
\toprule
& &  \multicolumn{7}{c}{Signal Strength ($\lambda$)} \\
\cmidrule{3-9}
Model & Method  &
 $0.15$ & $0.1$ & $0.07$ & $0.05$ & $0.03$ & $0.01$ & 
\multicolumn{1}{c@{}}{$0$} \\
\midrule
{Supervised LRT} & Asymptotic &   100 & 100 & 96 & 62 & 18 & 18 & 6\\
                     & Bootstrap &   100 & 96 & 78 & 58 & 6 & 0 & 0\\
				    & Permutation &  100 & 98 & 98 & 86 & 28 & 6 & 0 \\[6pt]
{Supervised Score} & Bootstrap &   64 & 66 & 74 & 50 & 18 & 0 & 0\\
                     & Permutation &   94 & 92 & 100 & 92 & 80 & 24 & 12 \\[6pt]
{Semi-Supervised LRT} & Asymptotic &  100 & 98 & 74 & 38 & 16 & 6 & 2\\
                     & Bootstrap  &   100 & 98 & 48 & 10 & 2 & 2 & 0\\
                     & Permutation &    100 & 98 & 72 & 38 & 16 & 6 & 2\\
                     & Slow Perm &   82 & 8 & 0 & 4 & 2 & 0 & 4\\[6pt]
{Semi-Supervised AUC} & Asymptotic &   100 & 96 & 78 & 32 & 14 & 4 & 2\\
                                       & Bootstrap &   100 & 98 & 70 & 32 & 20 & 6 & 2 \\
                                      & Permutation  &   100 & 98 & 68 & 32 & 20 & 4 & 2\\
                                      & Slow Perm &   100 & 100 & 94 & 56 & 20 & 8 & 4\\[6pt]
{Semi-Supervised MCE} & Asymptotic &   100 & 92 & 60 & 28 & 14 & 2 & 2\\
                                       & Bootstrap &   100 & 96 & 52 & 28 & 16 & 6 & 4 \\
                                      & Permutation  &   100 & 96 & 52 & 30 & 14 & 6 & 6\\
                                      & Slow Perm &  100 & 98 & 86 & 58 & 16 & 6 & 2\\
\bottomrule
\end{tabular}
\end{table}

\begin{figure}[h]
\includegraphics[width=\linewidth]{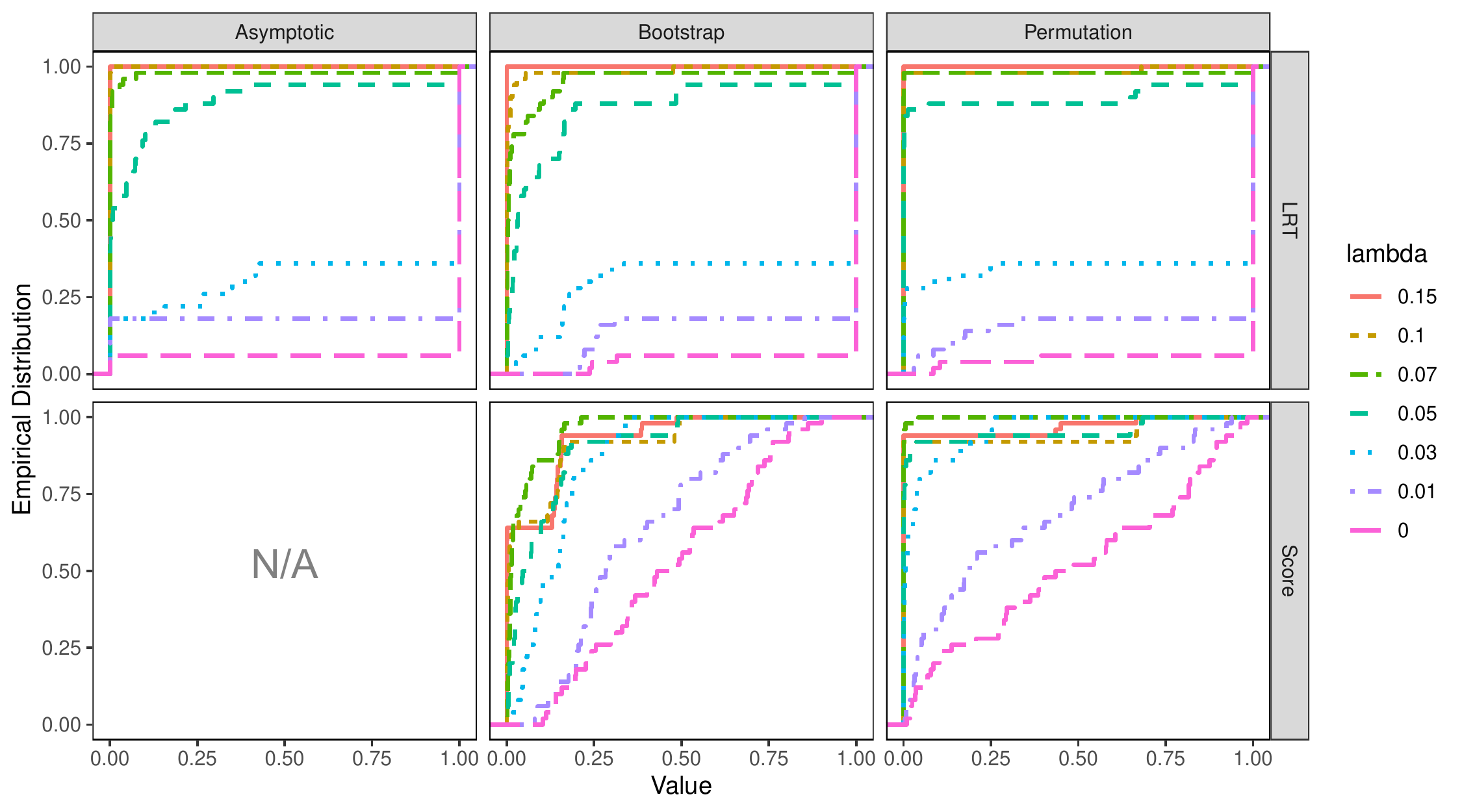}
\caption{Empirical CDF of the p-values given by the supervised tests for different signal strengths ($\lambda$). The columns in the grid of plots represent the different methods of testing (asymptotic, bootstrap and permutation) and the rows represent the use of the different test statistics (LRT and Score). Note that the plot for the asymptotic test using the score statistic is missing since we do not consider that test due to the poor quality of the asymptotic approximation.}
\label{fig:super}
\end{figure}

\begin{figure}[h]
\includegraphics[width=\linewidth]{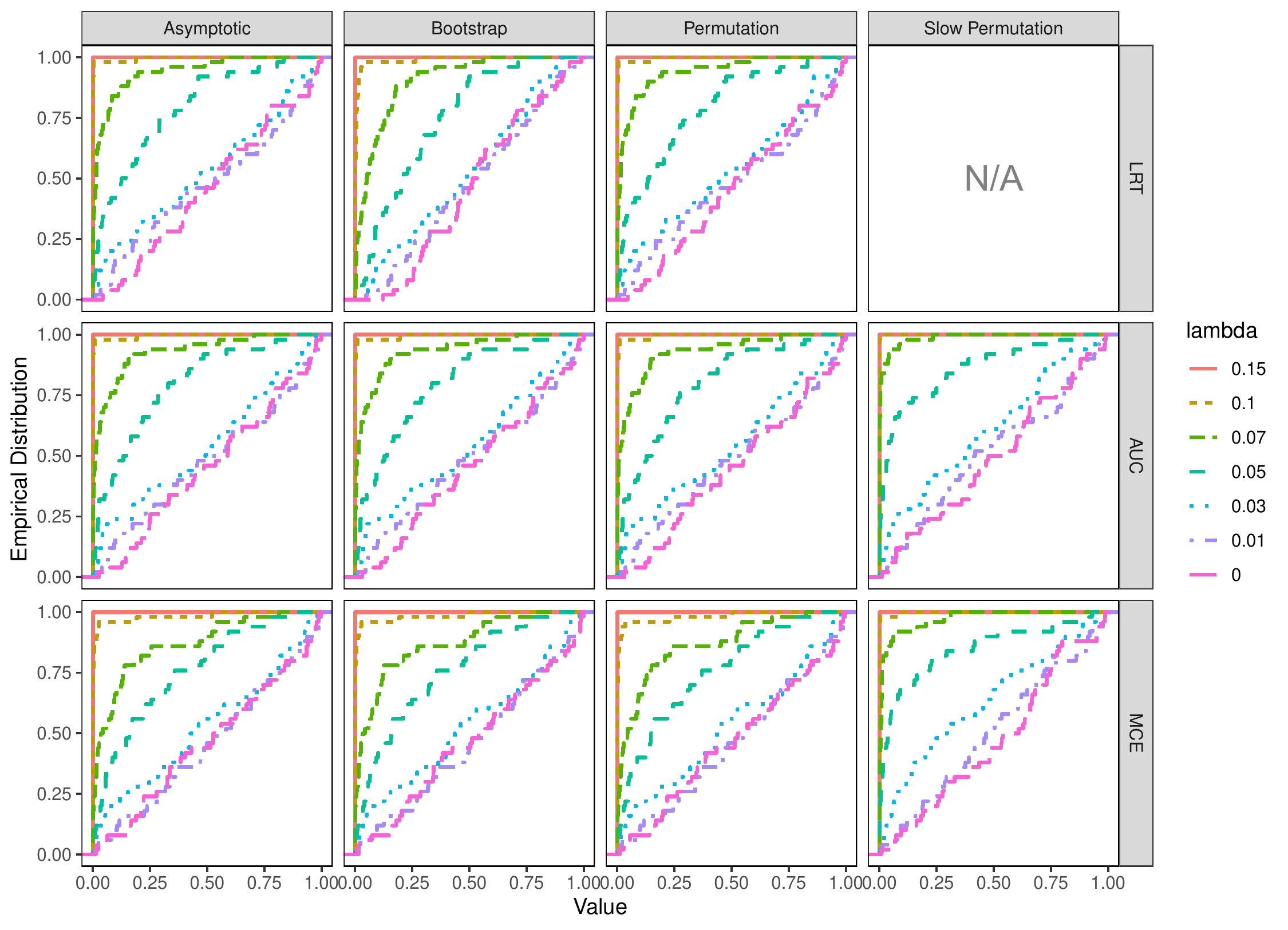}
\caption{Empirical CDF of the p-values given by the semi-supervised tests for different signal strengths ($\lambda$). The columns in the grid of plots represent the different methods of testing (asymptotic, bootstrap, permutation and slow permutation) and the rows represent the use of the different test statistics (LRT, AUC and MCE). We leave out the plot for slow permutation test using the LRT statistic since in practice it demonstrates very poor performance. It suffers from bias caused by over-fitting since for slow permutation we consider the in-sample LRT.}
\label{fig:semisuper}
\end{figure}

As seen in Table~\ref{tbl::Power}, among the supervised methods, the
permutation tests out-perform all the other supervised methods 
in terms of power. 
The permutation test with the score statistic 
has the most power for smaller values of $\lambda$.  
However, it is worth noting that it might be anticonservative 
for $\lambda = 0$ and hence, might not have the right significance level.
Conversely, as Table~\ref{tbl::Power} and Figure~\ref{fig:super} show, the supervised tests 
that use the LRT statistic appear to be overly conservative for 
smaller values of $\lambda$, because the LRT statistic uses an estimate
of the density ratio $p_s / p_b$ using the classifier.
This problem mainly appears when two conditions are both in effect: 
(1) the classifier for background versus signal overfits the class probabilities, by overfitting to the data, and (2) when $\lambda = 0$; i.e., when the experimental data has no signal.
Under this situaton, then, with probability higher than expected, 
almost the entire experimental data 
is classified as background with $\hat h({Z_i}) = 0$,
and $\widehat{p_s/p_b} = 0$. 
This makes $\hat{\lambda}_{\text{MLE}}$, 
the maximum likelihood estimator of $\lambda$, 
also zero, 
making the LRT statistic zero as well.
This is also observed when 
we use the permutation or bootstrap methods
to estimate the null distribution.
The estimated distributions have a
point mass at zero that
is much higher than $0.5$ 
(as given by the asymptotic distribution 
in Equation~(\ref{eqn::superLambda}) when true $p_s/p_b$ is used). 
Note that we additionally leave out the asymptotic score test 
both in Table~\ref{tbl::Power} and Figure~\ref{fig:super} 
since we tried the test for a sub-sample of the data 
and decided to not consider that test for the larger data set 
due to the poor quality of the asymptotic approximation.

Among the semi-supervised approaches, the AUC and MCE slow permutation methods (slow perm) have power that is only slightly worse than that of the supervised methods, as shown in Table~\ref{tbl::Power}. Importantly, the semi-supervised methods achieve this without having access to the labeled signal sample during training. Among the asymptotic and the faster permutation methods, using LRT or MCE gives similar performance to AUC in the semi-supervised approaches. We also observe that the slow permutation method using the LRT statistic has very low power. This is due to bias in estimating the LRT using an in-sample estimate which is influenced by over-fitting. 

Figure~\ref{fig:semisuper} shows the empirical distributions of the $p$-values produced by the semi-supervised tests for different $\lambda$ values. All the tests appear to have correct (or at least almost correct) type I error control as indicated by the close to uniform CDFs in the $\lambda = 0$ case. We also notice that none of the tests have much power to detect signals that are less than $3\%$ of the experimental data $(\lambda < 0.03)$. We omit the plot for the slow permutation method using the LRT as it has anomalously low power for most $\lambda$ values due to the reasons mentioned above.

We additionally compared these methods 
to nearest neighbor (NN) two-sample tests as
introduced in \cite{schilling1986multivariate} and
\cite{henze1988multivariate}. 
We considered the nearest neighbor tests 
as opposed to other tests,
since the signal that we are trying to detect,
appears as a collection of nearby data points. 
Hence tests based on neighborhoods 
should have better power to detect it.
We compared the methods presented in this paper 
to the NN tests (asymptotic and permutation versions) 
for a sub-sample of the data set. 
We observed that the asymptotic version especially had very poor power.
The permutation version performed better, but was
out-performed by the semi-supervised AUC, MCE and LRT methods.
Additionally, it was not scalable to extend it to the larger data set.
Hence we concluded that it was computationally impractical 
to apply it to the larger full data set.

In conclusion, we see that slow permutation method when using the AUC and MCE statistics out-performs the other semi-supervised methods and additionally gives comparable performance to the supervised methods in detecting the signal in the experimental data. 
Note that the power of the slow permutation method 
using the AUC and MCE statistics 
is much better than the one using the LRT statistic, 
which is the statistic used by \cite{d2019learning} and \cite{d2021learning}. 
So these methods give an improvement over using just the LRT statistic.

\subsubsection{Anomaly Detection when Data has a Misspecified Signal}

As mentioned in the introduction, 
a model-dependent search that targets one kind of new physics signal 
will not be powerful to detect a different signal which might actually be present in the data. 
This is one of the main motivations for model-independent methods. 
In this section, we demonstrate that if the signal model is misspecified in just one dimension, 
model-independent methods are still able to detect the signal, whereas the model-dependent methods fail to detect the signal.

We transform the signal in the experimental data 
to intentionally make it different from the signal model, 
which makes the signal model misspecified. 
We consider a transformation 
in just one variable, transforming 
\begin{equation}
\verb!tau_pt!^* = \verb!tau_pt! - 0.7 \ \left(\verb!tau_pt! - \min(\verb!tau_pt!) \right).
\end{equation}
in the experimental data. 
Meanwhile, we do not transform the signal data 
used by the supervised methods for training the classifier.
This simulates a misspecified signal situation, 
where the signal in the experimental data 
is not from the same distribution as the signal 
used by the supervised methods for training. 
This is to emulate a situation where the signal in the experimental data 
is different from the signal that the signal model specifies in just one dimension.

We chose this particular transformation for multiple reasons.
First, it transforms the variable that has the 
most marginally different means for the background 
and the signal data as demonstrated by Figure~4
in the Supplementary Material \citep{supplementary}. 
Second, the marginal distribution of the signal along this variable 
is different from the marginal distribution of the background.
These two reasons make \verb!tau_pt! a variable 
that potentially influences the classifier.
By transforming the variable, we lower the mean \verb!tau_pt!
for the signal in the experimental data, 
causing the model dependent methods to lose power  
in detecting the transformed signal.

We now compare the power of the 
supervised and the semi-supervised methods 
in detecting the transformed signal in the 
experimental data. 
The tests are run on $50$ random samplings of the data, 
and the percentage of times each of the tests rejects the null that there is no signal, 
is given in Table~\ref{tbl::PowerMiss}. 
As before, 
``Permutation" indicates the faster permutation method from Section~\ref{sec::semisupervised}, 
that uses out-of-sample test statistics for testing 
and ``Slow Perm" indicates the slower permutation method from Section~\ref{sec::semisupervised}, 
that uses in-sample test statistics for testing 
and re-trains the classifier in every permutation cycle. 
The significance level for all the tests is considered at $\alpha = 0.05$. 

\begin{table}[h]
\caption{Power of detecting the misspecified signal in the Higgs boson data for each model  in percentages. We consider 50 random samplings of the data and 1,000 bootstrap and permutation cycles. We perform each test at $\alpha = 5\%$ significance level. The last column ($\lambda = 0$) represents the Type I error of the methods.}
\label{tbl::PowerMiss}
\begin{tabular}{@{}lcrrrrrrr@{}}
\toprule
& &  \multicolumn{7}{c}{Signal Strength ($\lambda$)} \\
\cline{3-9}
Model & Method & 
 $0.15$ & $0.1$ & $0.07$ & $0.05$ & $0.03$ & $0.01$ & 
\multicolumn{1}{c@{}}{$0$} \\
\midrule
{Supervised LRT} & Asymptotic  &  2 & 10 & 2 & 8 & 8 & 6 & 4\\
                     & Bootstrap &   0 & 0 & 0 & 0 & 0 & 0 & 0\\
				    & Permutation &  0 & 0 & 0 & 0 & 0 & 2 & 0  \\[6pt]
{Supervised Score} & Bootstrap &  0 & 0 & 0 & 0 & 0 & 0 & 0\\
                     & Permutation & 0 & 0 & 0 & 0 & 0 & 2 & 8 \\[6pt]
{Semi-Supervised LRT} & Asymptotic &  100 & 100 & 100 & 82 & 18 & 4 & 4 \\
                     & Bootstrap  &  100 & 100 & 100 & 60 & 4 & 2 & 0\\
                     & Permutation &   100 & 100 & 100 & 82 & 18 & 4 & 2\\
                     & Slow Perm &  100 & 100 & 78 & 22 & 2 & 4 & 6 \\[6pt]
{Semi-Supervised AUC} & Asymptotic &  100 & 100 & 100 & 78 & 16 & 8 & 4\\
                                       & Bootstrap &  100 & 100 & 100 & 82 & 20 & 10 & 0 \\
                                      & Permutation &  100 & 100 & 100 & 80 & 20 & 8 & 2\\
                                      & Slow Perm &  100 & 100 & 100 & 100 & 34 & 10 & 4\\[6pt]
 {Semi-Supervised MCE} & Asymptotic &  100 & 100 & 100 & 66 & 24 & 6 & 4\\
                                       & Bootstrap &  100 & 100 & 100 & 62 & 16 & 6 & 4 \\
                                      & Permutation &  100 & 100 & 100 & 62 & 14 & 6 & 4\\
                                      & Slow Perm &  100 & 100 & 100 & 98 & 22 & 8 & 2\\       
\bottomrule
\end{tabular}
\end{table}

As seen in Table~\ref{tbl::PowerMiss},
as expected, the supervised methods 
have no power at all in detecting the transformed signal.
An interesting observation from the table is that, it appears that 
the supervised methods are more conservative 
when there is some signal present 
compared to when there is no signal present.
This occurs as the signal labels 
that the supervised models are trained on
are inconsistent with the signal in the experimental data, 
when the signal is present.

The semi-supervised methods on the other hand,
still have power to detect the transformed signal
since they are not trained on the misspecified signal training data. 
In fact, the semi-supervised methods appear to have higher power with the transformed signal than with the original signal, indicating that the transformed signal is somewhat easier to disentangle from the background than the original signal. 
Comparing the semi-supervised methods, 
we again observe that the slow permutation method using AUC has the highest power.
We also notice that the asymptotic and permutation tests 
that use LRT demonstrate power comparable to the AUC.
The slow permutation method with MCE also performs comparably to 
the slow permutation method with AUC.
The slow permutation method with LRT has again anomalously low power 
for smaller $\lambda$,
due to the bias caused by 
using the in-sample LRT as the test statistic.

In conclusion, semi-supervised methods continue to demonstrate power
similar to the previous case where the signal model was not misspecified.
On the other hand, the supervised models do not have any power
to detect the transformed signal in the misspecified model case,
which motivates the use of semi-supervised models in such a case.

\subsubsection{Anomaly Detection with Smaller Signal Strengths}
\label{sec::smalllambda}

So far, the signal strengths that we have considered in this paper have been in the order of $10^{-2}$ and the sample sizes have been in the order of $10^{4}$.  But, in many high-energy physics searches, the signal strengths that the physicists are looking for are in the order of $10^{-3} - 10^{-4}$ and they have much larger sample sizes as well. For example, the true signal strength of the Higgs Boson signal in the data set  where the two jets are produced is $\lambda = 0.0035$. In this section, we show that the power of the model-independent methods to detect signal for smaller signal strengths increases as the sample size increases.

We perform the asymptotic model-independent tests on equally sized background and experimental data sets whose sample sizes vary from $2 \times 10^4$ to $2 \times 10^6$, with signal strengths $\lambda$ varying from $0.001 - 0.05$. 
To make this experiment feasible (due to the lack of availability of sufficient background samples in our original data set), we fit a mixture of 8 multivariate Normals to the background data and a mixture of 9 multivariate Normals to the signal data in the Higgs boson data set. We then generate samples from the estimated mixtures of Normals to create our simulated data sets for this experiment.

\begin{figure}[t]
\includegraphics[width=\linewidth]{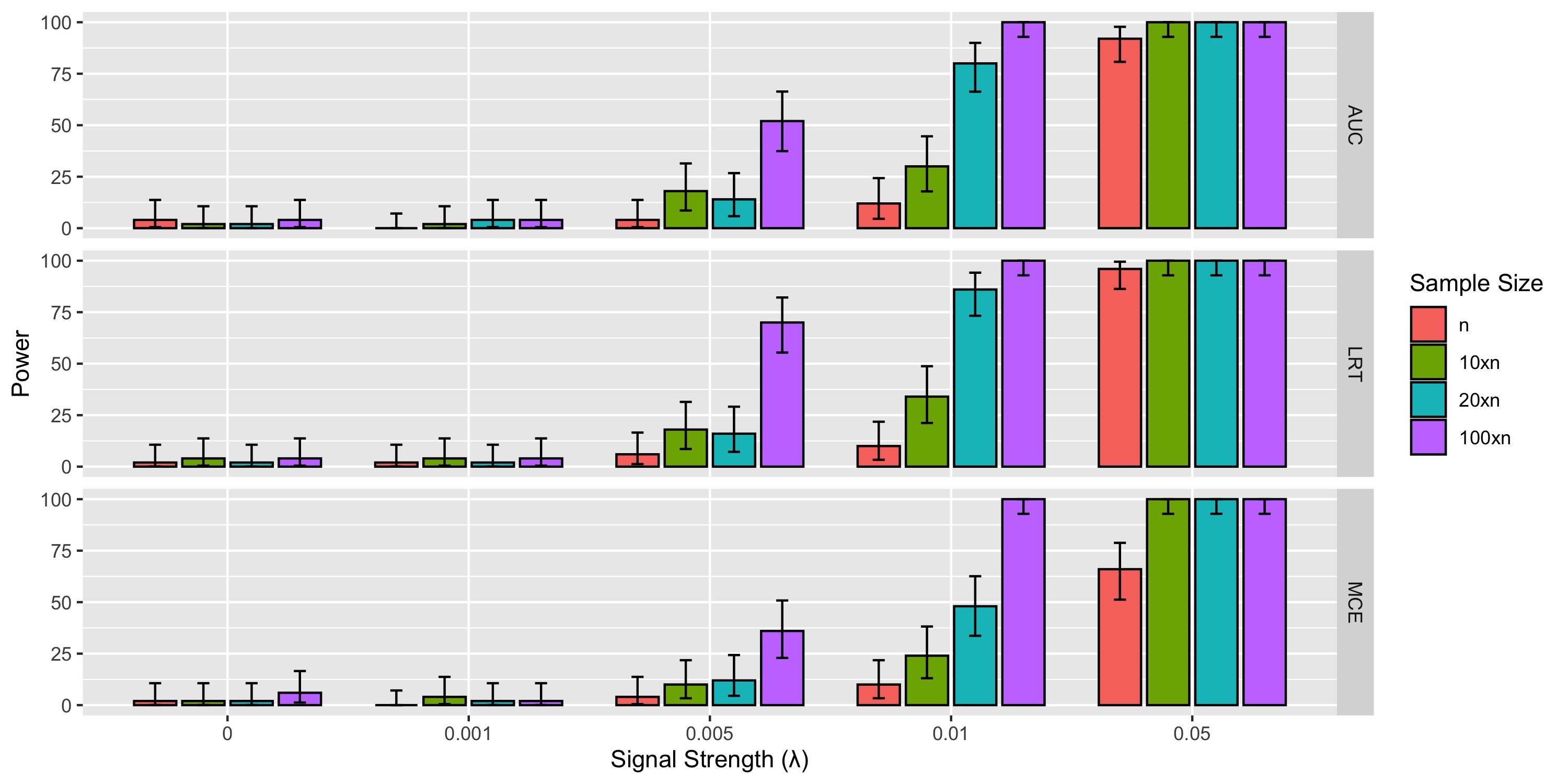}
\caption{Power of the asymptotic model-independent tests using  AUC, LRT, and MCE statistics to detect different signal strengths ($\lambda = 0, 0.001, 0.005, 0.01, 0.05$) in 50 simulations with samples of sizes $n, 10n, 20n, $ and $100n$, where $n = 2 \times 10^4$. $\lambda = 0$ represents the Type I error rate. The error bars represent the $95\%$ Clopper-Pearson confidence intervals for the power. We note that for all the three statistics, the power increases in larger sized samples for $\lambda = 0.005, 0.01, 0.05$. For $\lambda = 0.001$, none of the tests appear to have power to detect the signal, but further increases in the sample size should provide power in this case as well.}
\label{fig:powerplot}
\end{figure}

Figure~\ref{fig:powerplot} shows that for smaller signal strengths, for, e.g., $\lambda = 0.005$, the power of the model-independent tests increases as the sample size increases. We see that for the largest sample size, $100n = 2 \times 10^6$, the model-independent tests have good power even for $\lambda = 0.005$, which is comparable to the true Higgs Boson signal strength ($0.0035$) in the two jets channel.  Whereas, for the sample size that we used in our previous experiments, $n = 2 \times 10^4$, the model-independent tests have small power for $\lambda = 0.01$ and no power for $\lambda = 0.005$. This demonstrates that when larger data sets are available, the model-independent methods have power to detect even smaller signal strengths. In many search channels, LHC data sets are of the order of millions or billions of events, so these tests should have power for detecting reasonably small signals in those channels.

\subsection{Estimating the Higgs Boson Signal Strength}

In this section, we demonstrate the performance of the methods 
proposed in Section~\ref{sec::lambdaEst} 
to estimate the signal strength ($\lambda$) 
using the Higgs boson data set. 
We vary the signal strength from $\lambda = 0.5$ to
$\lambda = 0$ and look at the estimates of the signal strength
as well as the bootstrapped uncertainty intervals 
given by Method~\ref{mthd::est} in Section~\ref{sec::lambdaEst}.

We consider $100$ bootstrap cycles to 
get the $95\%$ bootstrapped uncertainty intervals.
For the estimates, we consider 
$T \in \{0.8, 0.5\}$ and $b \in \{0.01, 0.005\}$, 
where $T$ is the 
threshold that indicates the neighborhood of $1$
where we fit the Poisson regression model and 
$b$ is the bin-width of the histogram 
that we use to estimate the densities.
Note that bin-width $b \in \{0.01, 0.005\}$ is equivalent to
the number of bins, Bins $\in \{100, 200\}$.
To construct the bootstrapped uncertainty intervals,
we use three different methods using $\alpha = 5\%$ 
and compare them in 
Figure~\ref{fig:lambdaEst} below.

First, we use the basic bootstrap confidence interval,
also known as the Reverse Percentile Interval, 
which uses empirical quantiles 
of the bootstrap distribution of $\hat{\lambda}$ 
in a reverse order to construct the confidence interval 
(see \citet{davison1997bootstrap}, Eq. (5.6), p. 194):
$(2{\widehat {\lambda \,}}-\lambda _{(1-\alpha /2)}^{*},2{\widehat {\lambda \,}}-\lambda _{(\alpha /2)}^{*})$, where $\lambda_{(1-\alpha /2)}^{*}$ denotes the $1-\alpha /2$ quantile of the bootstrapped estimates $\lambda^{*}$.

Second, we use the bootstrapped quantiles to form 
percentile bootstrap confidence intervals 
(see  \citet{davison1997bootstrap}, Eq. (5.18), p. 203, 
\citet{efron1994introduction}, Eq. (13.5), p. 171):
$(\lambda _{(\alpha /2)}^{*},\lambda _{(1-\alpha /2)}^{*})$.

Third, we use bootstrapped standard errors 
and normal distribution's quantiles to create 
bootstrap confidence intervals:
$ (\widehat {\lambda} - z_{1 - \alpha /2} \ {\widehat {\text{se}}}_{\lambda^{*} },\widehat {\lambda} + z_{1 - \alpha /2} \ {\widehat {\text{se}}}_{\lambda^{*} })$ where $z_{1 - \alpha /2}$ denotes the $1 - \alpha /2$ quantile of the standard normal distribution, 
and ${\widehat {\text{se}}}_{\lambda^{*}}$ is the 
standard error estimated 
using the empirical standard deviation 
of the bootstrapped estimates $\lambda^{*}$.

We additionally present the confidence intervals
given by the GLM model \citep{dobson2018introduction} itself
in Figure~\ref{fig:lambdaEst}.

\begin{figure}[t]
\includegraphics[width=\linewidth]{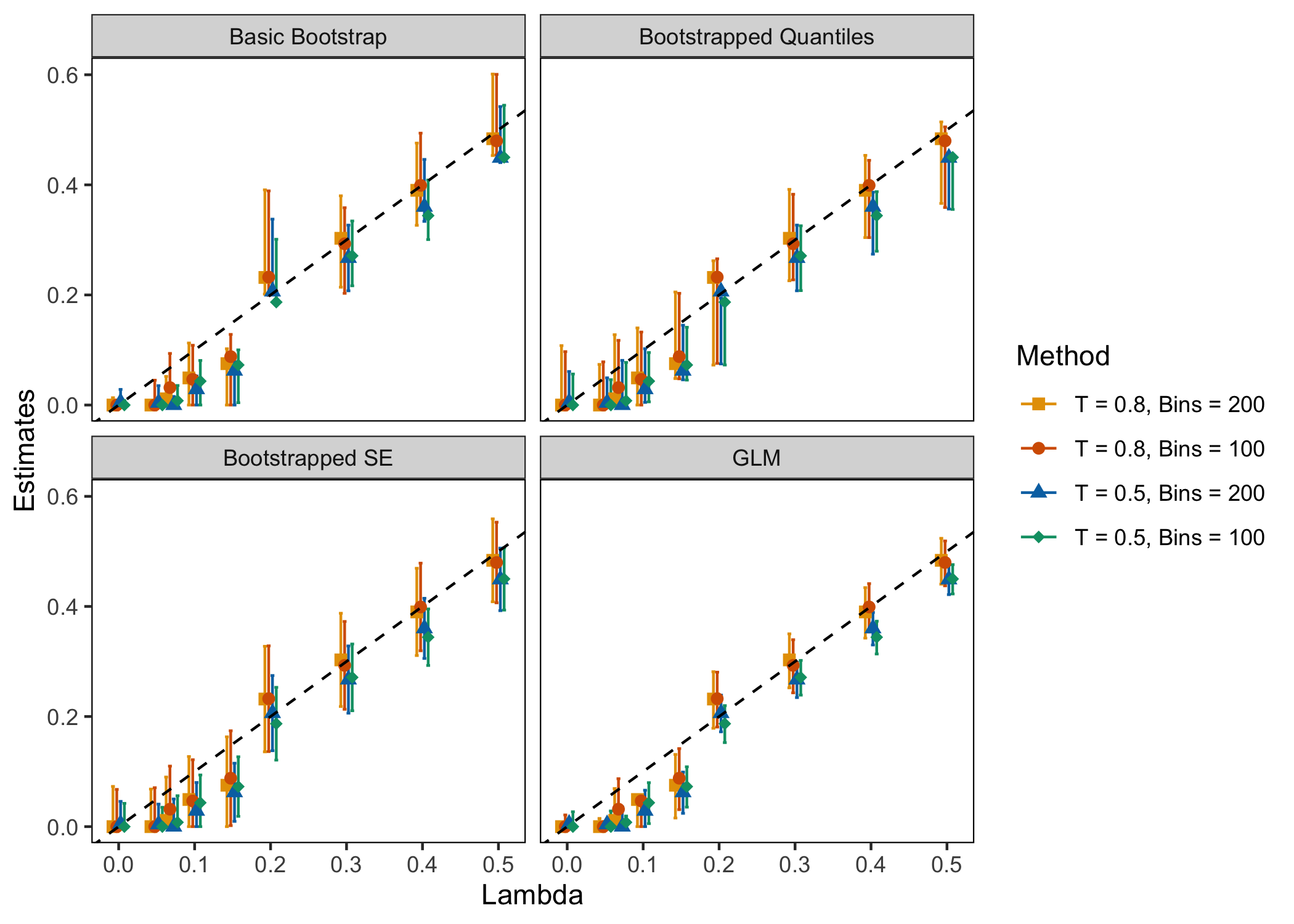}
\caption{Estimates of the signal strength ($\lambda$), along with the $95\%$ uncertainty intervals, of the Higgs boson in the experimental data. The true $\lambda$, given by the dotted diagonal line, varies from $0$ to $0.5$. $T$ specifies the threshold that controls the size of the neighborhood of $1$ that is considered in the Poisson regression model and Bins specifies the number of bins considered in the histogram when estimating the density of $\widehat{g_q}(1)$ as described in Section~\ref{sec::lambdaEst}.}
\label{fig:lambdaEst}
\end{figure}

We observe from Figure~\ref{fig:lambdaEst} 
that the estimates of $\lambda$ 
do not vary much with the number of bins used 
for the histogram estimate of the density. 
We also notice that thresholding at $T = 0.8$ 
which is closer to $1$ as compared to $T = 0.5$,
gives better estimates of $\lambda$. 
We additionally tried $T = 0.9$ on a sub-sample of the data set.
But the estimates in that case were not very stable and 
hence, we decided to not consider them for the final larger data set.
The uncertainty intervals created using 
bootstrapped quantiles and bootstrapped standard errors 
appear to be relatively well-calibrated when $T = 0.8$, i.e., 
they include the true value of $\lambda$, 
given by the dotted diagonal line, in each of the plots. 
The uncertainty intervals given by the GLM model and basic bootstrap, 
on the other hand,  
do not contain the true $\lambda$ for some smaller values of $\lambda$.
So, $\hat\lambda$ using $T = 0.8$ with the number of bins either $100$ or $200$,
accompanied with either 
bootstrapped quantile or bootstrapped standard error 
uncertainty intervals, 
appears to give the best performance in 
estimating and quantifying the uncertainty of the signal strength $\lambda$. 
An additional advantage of the uncertainty intervals using 
bootstrapped standard errors is that 
they are centered about the estimate $\hat\lambda$.

\subsection{Interpreting the Classifier Using Active Subspace Methods}
\label{sec::activeexamples}

We demonstrate the application of the active subspace methods for a
random simulation (one of the 50 simulations in
Section~\ref{sec::testperformance}) which detects the signal at
significance level $\alpha = 0.05$, when $\lambda = 0.15$, i.e., $15\%$ of the
experimental data is from the signal sample. We consider $\lambda =
0.15$, since the random forests demonstrate good power in detecting the
signal for that signal strength (Table~\ref{tbl::Power}).

We then use Method~\ref{mthd::active} presented in Section~\ref{sec::active} to find the active subspace of the fitted semi-supervised classifier. The second step of the algorithm requires us to choose a linear smoother as well as a smoothing parameter for it. We choose a Gaussian kernel smoother as the linear smoother and $h = 0.5$ as the smoothing parameter. The smoothing parameter $h$ is used to scale the standard deviations of the variables, which is then used as the standard deviation of the multivariate Gaussian kernel, i.e., $\hat{sd}(Z)/h$ is considered as the standard deviation of the multivariate Gaussian kernel. The smoothing parameter selection process is described in the Supplementary Material \citep{supplementary}.

\begin{figure}[H]
\begin{subfigure}{.49\textwidth}
  \centering
   \includegraphics[width=\linewidth]{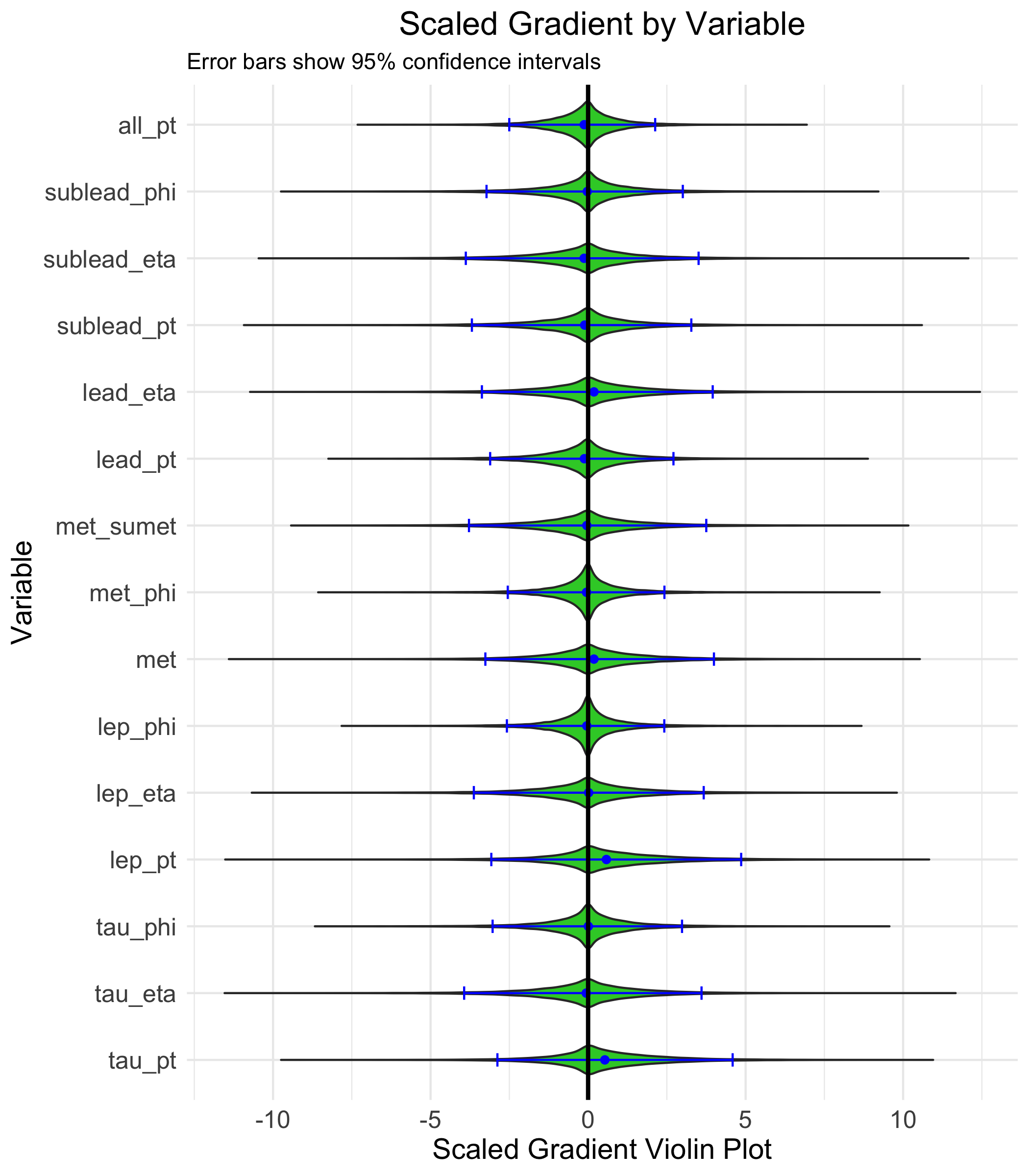}
  \caption{Standardized Gradient Distribution}
  \label{fig:meangrad}
\end{subfigure}
\begin{subfigure}{.49\textwidth}
  \centering
   \includegraphics[width=\linewidth]{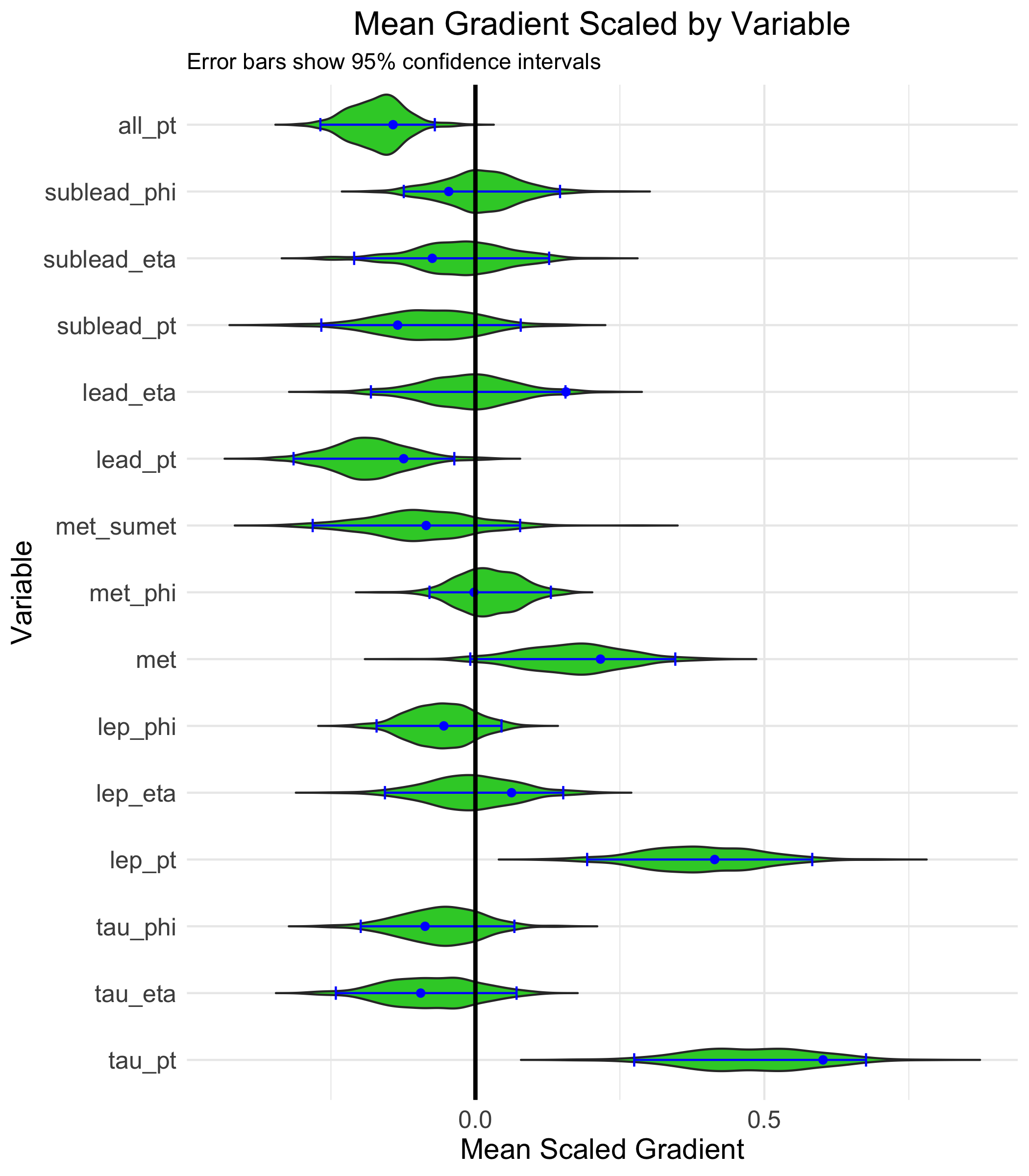}
  \caption{Mean Standardized Gradient}
  \label{fig:meangrad}
\end{subfigure}
\begin{subfigure}{.49\textwidth}
  \centering
   \includegraphics[width=\linewidth]{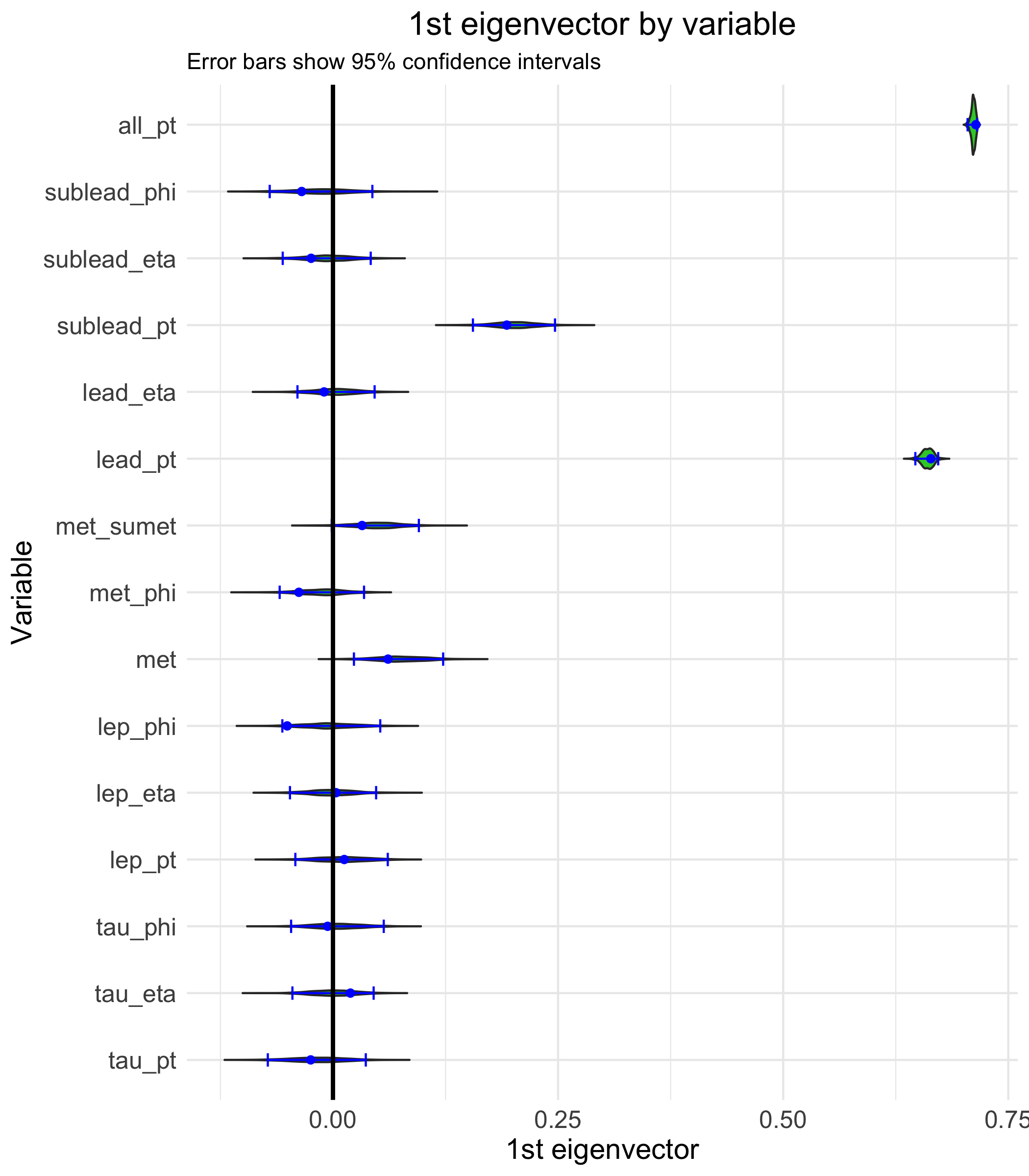}
  \caption{First Eigenvector}
  \label{fig:1eigen}
\end{subfigure}
\begin{subfigure}{.49\textwidth}
  \centering
   \includegraphics[width=\linewidth]{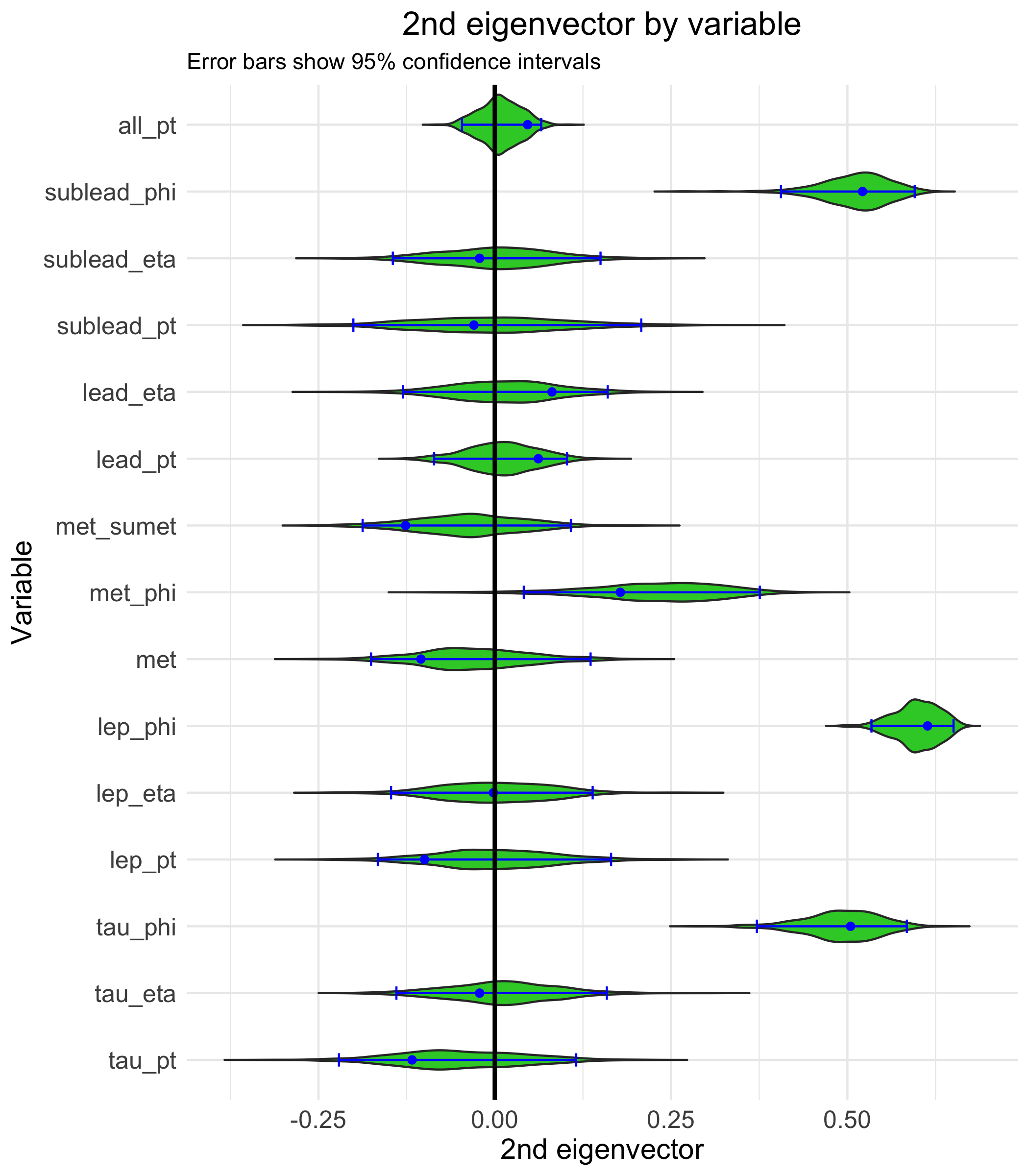}
  \caption{Second Eigenvector}
  \label{fig:2eigen}
\end{subfigure}
 \caption{The active subspace variables for the classifier trained on data with signal strength $\lambda = 0.15$ computed using a local linear smoother that uses the gaussian kernel with smoothing parameter $h = 0.5$. (a) gives the distribution of the standardized gradients $\hat{\beta}_j(Z_i)/\widehat{sd(\hat{\beta}_j(Z_i))}$ given by the local linear smoother at every point in the combined test data $Z_i \in \mathcal{X}_2 \cup \mathcal{W}_2$.  The dots denote the mean. In (b), (c) and (d), the violin plot and the dashes give the bootstrapped empirical distribution and the bootstrapped uncertainty intervals computed using the empirical quantiles respectively for the standardized mean, the first eigenvector and the second eigenvector. In (b) the dots give the mean standardized gradient similar to (a). In (c) and (d) the dots represent the first eigenvector and the second eigenvector computed on the combined test data, respectively.}
 \label{fig::activevariables}
\end{figure}

Figure~\ref{fig::activevariables} gives us the active variables given by the mean standardized gradient, the first eigenvector and the second eigenvector. The higher eigenvectors do not contain much information and have been included in the Supplementary Material \citep{supplementary}. The figure additionally gives the distribution of the bootstrapped active subspace estimates as violin plots. 
These help us construct $95\%$ uncertainty intervals 
for the active subspace variables using the bootstrapped empirical percentiles. 
We consider $500$ bootstrap cycles to 
get the $95\%$ bootstrapped uncertainty intervals.

Figure~\ref{fig::activevariables}(a) additionally provides the
violin plots of the standardized gradient estimates $\hat{\beta}_j(Z_i)/\widehat{sd(\hat{\beta}_j(Z_i))}$ given by the local linear smoother at every point in the combined test data $Z_i \in \mathcal{X}_2 \cup \mathcal{W}_2$. We notice that the violin plots are not very informative since they are all symmetric about zero, i.e., they don't appear to give any information about the slope of the classifier surface.

The mean standardized gradient in Figure~\ref{fig::activevariables}(b)
gives the direction in which the classifier output changes most rapidly on average. 
We see that jointly higher transverse momentums 
of the hadronic tau (\verb!tau_pt!) and the remaining lepton (\verb!lep_pt!)
contribute the most to an increase in the classifier output.
Increase in these transverse momentums,
combined with a decrease in the 
transverse momentum of the leading jet (\verb!lead_pt!) 
and the scalar sum of the transverse momentum of all the jets (\verb!all_pt!), 
leads to an increase in the classifier output.
This implies that the detected signal events display 
higher momentums of the hadronic tau and the other lepton, 
lower momentums of the leading jet and 
lower scalar sum of the momentums of all the jets 
as compared to background events. 
Note that the mean gradient vector 
not only captures the dependency of the classifier 
on each variable individually, 
but also characterizes the multivariable dependencies that influence the classifier.

The first eigenvector gives the 
first principal component of the gradients, 
which demonstrates the relationship between the variables 
that causes the most variability in the gradients of the classifier. 
The first eigenvector, as seen in Figure~\ref{fig::activevariables}(c), 
indicates that when \verb!all_pt!, the scalar sum of the transverse momentums, 
\verb!lead_pt!, the transverse momentum of the leading jet, 
\verb!sublead_pt!, the transverse momentum of the subleading jet  and 
\verb!met!, the missing transverse energy, 
jointly change in the same direction, 
it causes the most variation in the classifier. 
This means that these variables 
together, help the classifier separate the signal from the background.  
The second eigenvector, i.e., 
the second principal component of the gradients, 
as seen in Figure~\ref{fig::activevariables}(d), 
appears to capture the relationship 
between the azimuth angle $\phi$ of the different objects in the event  
that most influences the classifier. 
Recall that the $\phi$ values of the objects have been transformed 
to denote the difference in the angle between each object and the leading jet.
So, the second eigenvector indicates that 
the $\phi$ angle between the leading jet and all the other four objects
(the subleading jet, the hadronic tau, the lepton and the missing transverse energy)  
influences the classifier. 
This has an appealing physical interpretation 
in that it indicates that the azimuthal orientation 
between the leading jet and the rest of the event 
is a useful feature for separating the signal from the background.

\begin{Remark}
Note that for any eigenvector $M_{\cdot j}$ of the standardized gradients of the trained classifier, $ - M_{\cdot j}$ is also an eigenvector. 
This causes the violin plots of the bootstrapped eigenvectors 
to be systematically symmetric about zero. 
To solve this problem, we first take the variable that has the largest 
absolute eigenvector value, i.e. $k_j = \arg\max_{i} |\hat{M}_{ij}|$. 
Then we fix the sign of the bootstrapped eigenvectors $\hat{M}_{\cdot j}^*$
such that the sign of the $k_j^{th}$ variable matches, i.e., $\text{sgn}(\hat{M}_{k_j j}^*) = \text{sgn}(\hat{M}_{k_j j})$. 
This process has been followed in 
Figures~\ref{fig::activevariables}(c) and \ref{fig::activevariables}(d) 
 to handle the problem.
\end{Remark}

So, the active subspace methods 
provide an algorithm to interpret the semi-supervised classifier 
once it has detected a signal in the experimental data.
The methods imply that the classifier that detects the Higgs boson 
is positively influenced by \verb!tau_pt! and \verb!lep_pt!, and 
negatively influenced by \verb!all_pt! and \verb!lead_pt!. 
Additionally, it is also influenced by joint changes in the 
values of \verb!all_pt!, \verb!lead_pt!, \verb!lsubead_pt! and \verb!lmet!. 
The classifier is also influenced by the difference between the 
azimuth angle $\phi$ of the leading jet and those of the other four objects, 
indicating that this might be an important feature for detecting the signal events. 
Finally, we note that without techniques like these it would have been very difficult to understand what kind of a signal the semi-supervised classifier has detected within the high-dimensional feature space.

\section{Conclusion}
\label{sec::conclusion}

In this paper, we studied model-independent anomaly detection tests
using semi-supervised classifiers, that can detect the presence of
signal events hidden within background events in high energy particle
physics data sets. Additionally, we proposed methods to 
estimate the signal strength and  to
identify the active subspace affecting the classifier most strongly, leading
to an understanding of the detected signal. 
We demonstrated the performance of the methods and 
also compared the proposed
tests with comparable model-dependent supervised methods 
as well as nearest neighbor two-sample tests on a 
data set related to the search for the
Higgs boson at the Large Hadron Collider.

We presented multiple model-independent methods that search for the signal
without assuming any knowledge of the signal model. 
By not assuming any signal model,
we retain the ability to detect unknown and unexpected signals. 
This is an important capability for future searches at the Large Hadron Collider, where model-dependent searches have so far not yielded evidence of physics beyond the Standard Model. 
We used
a semi-supervised classifier to distinguish the experimental data from
the background data and used the performance of the classifier to
perform a test to detect a significant difference between the two data
sets. We proposed three test statistics that can be used for the test: 
the likelihood
ratio test (LRT) statistic, the area under the curve 
(AUC)  statistic and the misclassification error (MCE) statistic. 

We compared 
the use of the different test statistics,  
as well as the use of different techniques for obtaining the null distribution, 
in building the test that has the most power in detecting the signal.
We compared the power of the methods to detect the Higgs boson at
different signal strengths and showed that a version of the proposed AUC
method has power that is competitive with the model-dependent methods. 
So, even when the signal model is correctly assumed by the model-dependent
methods, the proposed model-independent methods appear to still have competitive 
power to detect the presence of the signal. However, when the signal
model is incorrectly assumed or misspecified, 
the model-dependent methods can totally miss the signal, 
whereas the proposed model-independent ones are
still able to detect the signal, as demonstrated in our experiments. 
In particular, the proposed methods
demonstrate the ability to find new particles without specific a priori
knowledge of their properties.

As described in Section~\ref{sec::supervised}, 
model-dependent methods currently in use in experimental high-energy physics, 
are slightly different from the ones presented in Section~\ref{sec::supervised}. 
In this paper, to make the model-dependent methods comparable 
to the proposed model-independent ones, 
we consider high-dimensional model-dependent tests. 
In current practice though, a threshold is placed on the supervised classifier output 
to select a subset of the experimental data that is richer in signal events. 
Then the high-dimensional data set is transformed into a univariate one, and the transformed data are fitted using a one-dimensional mixture model consisting of signal and background components. This is used to construct a profile likelihood ratio test \citep{atlas2011collaborations}, where some of the nuisance parameters 
are related to the one-dimensional background model. 
\cite{dauncey2015handling} proposed a discrete profiling method for this test
based on considering the choice of the background functional form as a discrete
nuisance parameter which is 
profiled in an analogous way to continuous nuisance parameters.

These considerations motivate multiple avenues for future exploration. We could construct an analogous model-independent 
version of the approach described above, where we first use the semi-supervised classifier 
output to select a signal-rich subset of the experimental data, 
then transform the selected data to a one-dimensional space and 
finally perform a test in the one-dimensional space using a mixture model. For well-chosen transformations, this could increase the power of the test. More interestingly, for many signals, the signal distribution corresponding to a transformation into the invariant mass variable is predicted by quantum mechanics to be the Cauchy distribution (also known as the Breit--Wigner distribution in particle physics; see Chapter 49 in \cite{Zyla:2020zbs}), whose parameters, under the model-independent scenario, are unknown to us. It would be interesting to built this knowledge into the semi-supervised method. A straightforward approach would be to simply use the Cauchy distribution, convolved with a model for detector smearing, in the one-dimensional mixture model, but it might also be possible to incorporate this knowledge into the training of the high-dimensional semi-supervised classifier.

Another important avenue of future work would be to find ways to account for systematic uncertainties in the background training data. As described above, the current supervised techniques account for systematics using profile likelihoods for one-dimensional summary statistics, with the systematic variations in the one-dimensional space parameterized using potentially hundreds or thousands of nuisance parameters. This should also be feasible for one-dimensional semi-supervised tests. However, it might also be possible to use profiling in the high-dimensional semi-supervised likelihood ratio tests. For example, if we had two plausible background samples from two different MC generators, we could use optimal transport \citep{peyre2019computational} to find the geodesic between the samples. We could then profile over the parameter corresponding to the location on the geodesic, which would account for the systematic uncertainty corresponding to the envelope of background models spanned by the two MC generators.


\begin{acks}[Acknowledgments]
The authors would like to thank the anonymous referees, the Associate
Editor and the Editor of Annals of Applied Statistics for their extensive, thoughtful  and constructive comments that greatly improved the quality of this paper.
\end{acks}
\begin{funding}
This work was supported in part by NSF awards PHY-2020295, DMS-2053804 and DMS-2113684.
\end{funding}

\begin{supplement}
\stitle{Supplement to ``Model-Independent Detection of New Physics Signals Using Interpretable Semi-Supervised Classifier Tests''}
\sdescription{The supplementary material contains the proof of Theorem~\ref{thm::Lambda}, some of the proposed algorithms from Section~\ref{sec::semisupervised}, and details about the exploratory data analysis of the Higgs boson data used in the experiments in Section~\ref{sec::higgs}. It additionally describes the selection of the smoothing parameter for the active subspace methods. The supplementary material can be found at \url{https://github.com/purvashac/MIDetectionClassifierTests/blob/main/Supplementary.pdf}}
\end{supplement}


\bibliographystyle{imsart-nameyear} 
\bibliography{paper}       

\end{document}